\documentclass[final,5p,times]{elsarticle}

\usepackage{elsREP}

\RequirePackage{amsmath}

\RequirePackage{booktabs}

\RequirePackage{graphicx}

\RequirePackage[colorlinks=true,hyperfootnotes=false,bookmarksnumbered=true]{hyperref}
\RequirePackage[all]{hypcap}
\RequirePackage[capitalise,nameinlink,noabbrev]{cleveref}

\biboptions{sort&compress}
\journal{arXiv}

\newcommand{\Qi}{Q_{ext,1}}
\newcommand{\Qp}{Q_{ext,2}}
\newcommand{\Qos}{Q_{0s}}
\newcommand{\sto}{|S_{21}|}
\newcommand{\soo}{|S_{11}|}
\newcommand{\rto}{|R_{21}|}
\newcommand{\tten}[1]{\!\cdot\! 10^{#1}}

\renewcommand{\today}{\number\day\space \ifcase\month\or
  January\or February\or March\or April\or May\or June\or
  July\or August\or September\or October\or November\or December\fi
  \space \number\year}

\hypersetup{pdftitle={Two-port CW measurements on RF cavities: Notes on self-consistency assessment and indirect methods},
            pdfauthor={W. Hartung et al.}}

\begin{document}

\begin{frontmatter}

\title{Two-port CW measurements on RF cavities: Notes on self-consistency assessment and indirect methods}

\author{Walter H. Hartung}
\author{Wei Chang}
\author{Sang-Hoon Kim}
\author{Taro Konomi}
\author{Ting Xu}

\address{Facility for Rare Isotope Beams, Michigan State University, East Lansing, Michigan, USA}
\date{20 January 2026}

\begin{abstract}
In the case of a radio-frequency (RF) cavity with a mismatched input
coupler, a direct calculation of the power dissipation in the cavity
and the intrinsic quality factor from continuous-wave (CW) measurements
may have uncertainty due to systematic errors.  Formulae for an
indirect calculation of these quantities are derived for the case of
a cavity with two couplers of fixed coupling strength.  In this
approach, the signal from the pickup coupler is used to infer the
amplitude of the ``emitted wave'' from the input coupler.  A
graphical method for self-consistency assessment is evaluated.  The
impact of frequency offsets is considered.  Applications of
these methods are presented, drawing on cold tests of
superconducting cavities produced for the Facility for Rare Isotope
Beams.
\end{abstract}

\end{frontmatter}

\tableofcontents

\section{Introduction\label{S:intro}}

Continuous-wave (CW) measurements are used to evaluate the performance
of radio-frequency (RF) cavities.  RF cavities are used in charged
particle accelerators, among other applications.  The cavity stored
energy, field amplitude, accelerating gradient, and quality factor may
be inferred from CW measurements.  RF cavity testing is an important
part of cavity development, surface preparation development, and
cavity production for small- and large-scale accelerators.

For normal-conducting RF (NRF) cavities, single-port or multi-port
measurements are commonly used.  High-field testing may require a
high-power amplifier and water cooling of the cavity exterior.
Feedback may not be needed, as the bandwidth may be such that the
amplitude and phase are relatively stable with an open-loop RF system.

For superconducting RF (SRF) cavities, 2-port
measurements are typical.  The cavity must be cooled to cryogenic
temperatures, but usually the test can be done with a small
amplifier (50 to 300~W being typical).  Due to the narrow
bandwidth, a phase-lock loop or a self-excited loop is typically used
to ensure that the cavity is driven on resonance.

Methods for measurements on RF cavities and analysis of the results
can be found in the literature, for example in
Refs.~\cite{GINZTON1957, PKH1998, WANGLER2008}.  Typically the cavity
is modeled as an $LRC$ circuit; though both a circuit model and a
waveguide model for the cavity-coupler system are considered in
Chapter 5 of Ref.~\cite{WANGLER2008}.  Considerations specific to SRF
cavity case are discussed in Chapter 8 of Ref.~\cite{PKH1998}.

Systematic errors are generally present in RF measurements due to
imperfect components and imperfect matching.  The systematic errors
may produce large errors in the inferred quantities if the input
coupler is poorly matched to the cavity. A variable coupler may be
used to minimize the input coupler mismatch, but such a coupler makes
the system more complex, and there are sometimes concerns about
increasing the risk of particulate contamination due to the materials
and moving parts.

In the presence of systematic measurement errors, additional care is
needed in the analysis of RF measurements.  Methods to check the
consistency of derived quantities are useful in assessing the impact of
systematic errors and the validity of the results.  A thorough
understanding of systematic errors becomes particularly important in
cavity development efforts oriented toward performance improvements
with the need for precise performance comparisons between different
cavity designs, different surface treatments, and so on.  Some of
these issues are discussed in recent papers on SRF cavity
measurements, including an evaluation of the effect of an imperfect
circulator or an imperfect dual-directional coupler; an error analysis
for derived quantities; and error models with vector correction in
lieu of scalar correction \cite{NIMA830:22, NIMA913:7}.

In both the NRF and SRF cases, a basic analysis of CW measurements
may be done via a ``direct'' approach: the power dissipation in
the cavity is calculated from the difference between the forward power
supplied by the RF system and the reverse power from the input
coupler.  However, in the 2-port case with fixed coupling strengths, CW
measurements are over-determined, which allows for an ``indirect'' method
to calculate derived quantities and allows us to check the consistency
of the results.

This report describes the direct and indirect analysis methods, along
with a graphical method to check the self-consistency of the
measurements.  The methods can be applied to 2-port CW measurements
with fixed coupling strength for both NRF and SRF cavities.  A
weakly-coupled pickup coupler is assumed to simplify the derivation;
for completeness, the case of arbitrary pickup coupling strength is
considered in the appendices.  Vector measurements are not needed to
apply these methods.

The graphical methods described in this report have been applied to
SRF cavity tests at Michigan State University (MSU) for a number of
years.  Work at MSU has included SRF cavity development, prototyping,
and production for an ion re-accelerator linac \cite{LINAC2010:THP039}
and the driver linac for the Facility for Rare Isotope Beams (FRIB)
\cite{SRF2021:MOOFAV10}.  The latter effort required cold testing and
certification of 324 SRF cavities \cite{NIMA1014:165675}.
Certification testing is done with close-to-matched input coupling and
we generally use the direct method to infer the quality factor.
An additional effort in the past several years has been oriented
toward the development of multi-cell cavities for the proposed FRIB
energy upgrade \cite{NIMA1059:168985}.

In more recent years, the MSU SRF team has begun developing techniques
for in-situ plasma processing of FRIB cavities \cite{SRF2023:THIXA01}.
The plasma is driven with RF power via the fundamental power coupler
(FPC), but the cavity is at room temperature.  As a result, the FPC is
weakly coupled.  We have found that the indirect method is useful to
infer the quality factor and power dissipation with plasma on, and to
estimate the power transfer to the plasma \cite{SCM813:REPORT:ARXIV}.

This report describes the indirect and graphical methods in more
detail and includes some examples based on SRF cavity testing at
MSU\@.  Some of the derivations were previously included in an
appendix of a PhD dissertation by D. J. Meidlinger
\cite{DJMEIDLINGER2007}.  The present report includes a description of
the ``emitted wave'' picture, which provided the original inspiration
for the methods, but which was not previously documented.  Limiting
cases and the impact of frequency offsets are discussed as well.

\section{Background\label{S:back}}

\subsection{CW Measurement Basics\label{S:base}}

We may measure the following quantities in CW mode: the forward power 
($P_f$); the reverse power ($P_r$); the transmitted power
($P_t$); the resonant frequency ($f_0$).
These quantities are measured on resonance, with the drive frequency
set for maximum $P_t$ and minimum $P_r$.  Derived quantities of
interest include the stored energy ($U$), the accelerating gradient
($E_a$), the power dissipation in the cavity walls ($P_d$), and the
intrinsic quality factor
\begin{equation}
Q_0 \equiv \frac{\omega_0 U}{P_d}\, ,\label{eq:qo}
\end{equation}
where $\omega_0 = 2\pi f_0$ is the resonant angular frequency.  Other useful quantities are
the strength of the input coupler
($\Qi$) and the strength of the pickup coupler
($\Qp$).
The latter can be defined straightforwardly as
\begin{equation}
\Qp \equiv \frac{\omega_0 U}{P_t}\, .\label{eq:qp}
\end{equation}
Additional useful quantities are the coupling factors
\begin{align}
\beta_1 &\equiv \frac{Q_0}{\Qi}\label{eq:b1}\, ,\\
\beta_2 &\equiv \frac{Q_0}{\Qp}\label{eq:b2}\, .
\end{align}
Though we refer to $\Qi$ and $\Qp$ as coupling strengths and refer to
$\beta_1$ and $\beta_2$ as coupling factors, others may use the
opposite terminology.  Using \cref{eq:qo} and
\cref{eq:qp}, we can express $\beta_2$ as a power ratio:
\begin{equation}
\beta_2 = \frac{P_t}{P_d}\, .\label{eq:b2p}
\end{equation}

We define the ``scattering parameters''
\begin{align}
\soo &\equiv \sqrt{\frac{P_r}{P_f}}\label{eq:s11}\, ,\\
\sto &\equiv \sqrt{\frac{P_t}{P_f}}\label{eq:s21}\, ,
\end{align}
which are typically measured on warm cavities at low power using
a network analyzer, but can be obtained from power measurements as well.

Additional quantities of interest include the loaded quality factor
$Q_L$, which can be calculated via
\begin{equation}
\frac{1}{Q_L} = \frac{1}{Q_0} + \frac{1}{\Qi} + \frac{1}{\Qp}\label{eq:QLinv}
\end{equation}
and the loaded bandwidth
\begin{equation}
\Delta f \equiv \frac{f_0}{Q_L} \approx \frac{f_0}{Q_0} (1 + \beta_1)\, ,\label{eq:BW}
\end{equation}
with the right-hand side being an approximation for $\beta_2 \ll 1$.

\subsection{RF Power and Field Level Calibration\label{S:RFcal}}

The RF power ($P_f$, $P_r$, $P_t$) generally cannot be measured
directly.  Hence RF calibrations are needed to account for the effect
of directional couplers, cable attenuation, and other features of the
RF system.  This allows us to convert from the ``signal'' power values
to the ``actual'' power values at the input and pickup ports of the
cavity.  In this report, it is assumed that these calibrations have
been done, so that $P_f$, $P_r$, and $P_t$ are actual values.
More in-depth discussions of RF calibrations can be found elsewhere, for
example Ref. \cite{SRF2005:SUP02}.

CW measurements alone are not generally sufficient to calculate all of
the derived quantities of interest.  We can obtain additional
information by doing a ``field level calibration'' at low power.  For
SRF cavities, this is typically done with amplitude modulation.  An
overview of modulated measurement analysis is provided in \ref{S:Mod},
considering both the traditional case of weak input coupling and the
general case of arbitrary pickup coupling.  Both $\Qi$ and $\Qp$ can
be calculated from modulated measurements, along with $U$ and $Q_0$.
For NRF cavities, it is more practical to sweep the frequency rather
than modulate the amplitude; an equivalent field level calibration can
be done in the frequency domain.

A numerical calculation of the field pattern is needed to establish the
relationship between the stored energy and the fields for the mode of
interest.  This may be done using 2D codes such as SUPERFISH
\cite{PA7:213, PAC1993:790} or SuperLANS \cite{PAC1991:3002}, or 3D
codes such as CST Microwave Studio \cite{ICAP2006:THM2IS03} or Omega3P
\cite{ICAP2015:FRAJI3}.  The accelerating gradient $E_a$ is
proportional to the square root of the stored energy:
\begin{equation}
E_a \propto \sqrt{U}\, .
\label{eq:field}
\end{equation}
The peak surface electric field and peak surface magnetic field are
proportional to the square root of $U$ as well.  These and other
derived quantities such as the accelerating voltage may be calculated
from $U$ using the output of the numerical calculation.

\section{CW Analysis}

If $\Qi$ and $\Qp$ are known (generally from a low-power field
calibration, per \cref{S:RFcal}), the CW measurements provide more
information than is needed, such that there is more than one way to
calculate the stored energy, field, and $Q_0$.  In principle, the
different methods should give the same result.  In practice, the
answers may be different due to systematic errors, as described above.

\subsection{Direct Method Using \texorpdfstring{$P_f - P_r - P_t$}{Pf - Pr - Pt}\label{S:dir}}

The dissipated power $P_d$ is calculated by subtraction,
\begin{equation}
P_d = P_f - P_r - P_t\, .\label{eq:pdiff}
\end{equation}
The stored energy is related to the transmitted power via \cref{eq:qp}.
Knowing $P_t$, $\Qp$, and $f_0$, $U$ may be calculated by rewriting
\cref{eq:qp} as
\begin{equation}
U = P_t \frac{\Qp}{\omega_0}\, .\label{eq:Upt}
\end{equation}
With $f_0$, $P_d$, and $U$ known, $Q_0$ may be calculated directly via
\cref{eq:qo}.
In terms of measured RF power values, the intrinsic $Q$ is
\begin{equation}
Q_0 = \Qp\left(\frac{P_t}{P_f - P_r - P_t}\right)\, .\label{eq:Qdir}
\end{equation}
In terms of measured S-parameters, we have
\begin{equation}
Q_0 = \Qp\left(\frac{\sto^2}{1 - \soo^2 - \sto^2}\right)\, .\label{eq:QdirS}
\end{equation}
The field may be calculated from $U$ via \cref{eq:field}
with the appropriate scaling coefficient.

Advantages of the direct method include (i) it is straightforward to
understand and use; (ii) knowledge of $\Qi$ is not needed.  An
additional advantage is that the formulae derived above are valid for
arbitrary pickup coupler strength.

Disadvantages of the direct method include
(i) it makes use of $P_r$, which may have significant systematic error associated with it;
(ii) it relies on the difference between $P_f$ and $P_r$;
when the input coupling factor is not close to
unity, $P_f - P_r$ is a small number that is the difference of 2
large numbers, which therefore tends to magnify systematic errors.

\subsection{Indirect Method Using \texorpdfstring{$P_t/P_f$}{Pt/Pf}: Brute
  Force Derivation\label{S:ind}}

In the previous section, we used 3 equations, namely 
\cref{eq:pdiff}, \cref{eq:Upt}, and \cref{eq:qo}, to solve for 3
unknowns, namely $P_d$, $U$, and $Q_0$.  We would now like to derive
an expression for $Q_0$ which does not make use of $P_r$.  This will
require an additional equation.  We begin with
\cref{eq:pdiff}, re-writing it in terms of the scattering
parameters defined via \cref{eq:s11} and
\cref{eq:s21} to obtain
\begin{equation}
\frac{P_d}{P_f} = 1 - \soo^2-\sto^2\, .
\label{eq:normalPdiff}
\end{equation}
Our additional equation relates the input coupling factor $\beta_1$ to the ratio $P_r/P_f$
via \cite[Section 8.3]{PKH1998}:
\begin{equation}
\beta_1 \approx \frac{1 \pm \sqrt{P_r/P_f}}{1 \mp \sqrt{P_r/P_f}}\, .\label{eq:betaRat}
\end{equation}
In the ``$\pm$'' and ``$\mp$''operators, the upper sign applies for
the over-coupled case; the lower sign applies for the under-coupled
case.  The above equation is an approximation valid if
$\beta_2 \ll 1$.  In terms of $S$-parameters, we can write
\begin{equation}
\beta_1 \approx \frac{1\pm\soo}{1\mp\soo}\, .\label{eq:betaoS}
\end{equation}
Solving for $\soo^2$, we obtain
\begin{equation}
\soo^2 \approx \left(\frac{1-\beta_1}{1+\beta_1}\right)^2\, .
\end{equation}
Substituting this expression into \cref{eq:normalPdiff} and
simplifying gives
\begin{equation}
\frac{P_d}{P_f} \approx \frac{4\beta_1}{(1+\beta_1)^2}-\sto^2\, .
\label{eq:simple}
\end{equation}
Since $\beta_1$ and $P_d$ are both unknown, we need to eliminate
$P_d$.  Using \cref{eq:b2p} and \cref{eq:s21}, we can
write $P_d$ in terms of $P_f$, $\beta_2$, and $\sto$:
\begin{equation}
\frac{P_d}{P_f} \approx \frac{\sto^2}{\beta_2}\, .
\end{equation}
Substituting this into \cref{eq:simple} gives
\begin{equation}
\frac{\sto^2}{\beta_2} \approx \frac{4\beta_1}{(1+\beta_1)^2}-\sto^2\, .
\label{eq:s21Pdiff} 
\end{equation}
Let us define a normalized scattering parameter:
\begin{equation}
R_{21} \equiv S_{21} \sqrt{\frac{\beta_1}{\beta_2}} = S_{21} \sqrt{\frac{\Qp}{\Qi}}\, , 
\label{eq:define}
\end{equation}
so that, after multiplying by $\beta_1$ \cref{eq:s21Pdiff} becomes,
\begin{equation}
\rto^2 \approx \frac{4\beta_1^2}{(1+\beta_1)^2} - \beta_2\rto^2\, , 
\end{equation}
which can be rearranged to give
\begin{equation}
\rto\sqrt{1+\beta_2} \approx \frac{2\beta_1}{1+\beta_1}\, .
\end{equation}
If $\beta_2\ll 1$ (\emph{i.e.} weakly-coupled pickup antenna, as
assumed above), then 
\begin{equation}
\rto \approx \frac{2\beta_1}{1+\beta_1}\, .\label{eq:RtoB}
\end{equation}
Solving for $\beta_1$, we obtain
\begin{equation}
\beta_1 \approx \frac{1}{\frac{2}{\rto}-1} = \frac{\rto}{2 - \rto}\, .\label{eq:betaoR}
\end{equation}
Using \cref{eq:b1}, we can obtain an expression for $Q_0$,
\begin{equation}
Q_0 \approx \frac{\Qi}{\frac{2}{\rto}-1} = \left(\frac{\rto}{2 - \rto}\right) \Qi\, .\label{eq:qor}
\end{equation}
Substituting \cref{eq:define} and \cref{eq:s21} into the
last expression results in an explicit expression for $Q_0$ in terms
of measured powers and coupling strengths,
\begin{equation}
Q_0 \approx \frac{\Qi}{2\sqrt{\frac{\Qi}{\Qp}\cdot\frac{P_f}{P_t}}-1}\, .\label{eq:qzeropt}
\end{equation} 

\subsection{Indirect Method Using \texorpdfstring{$P_t/P_f$}{Pt/Pf}: Emitted
  Wave Derivation\label{S:emit}}

In Chapter 8 of Ref.~\cite{PKH1998}, the cavity behavior for cases other than the steady-state
case are considered.  If the drive power is switched off ($P_f \rightarrow
0$ at time $t = 0$), one can consider an ``emitted power'' ($P_e$) which
is the instantaneous value of $P_r$ after $P_f$ becomes zero.  This
quantity is related to the coupling strength via~\cite[Section 8.2.1]{PKH1998}
\begin{equation}
\Qi = \frac{\omega_0 U}{P_e}\, .\label{eq:qi}
\end{equation}
In this case, $U$ is the value of the stored energy at $t = 0$; the
stored energy will then decrease exponentially with time.

The input coupling factor can be
expressed in terms of $P_e$ via~\cite[Section 8.3.1]{PKH1998}
\begin{equation}
\beta_1 \approx \frac{1}{2 \sqrt{\frac{P_f}{P_e}} - 1}\, .\label{eq:bpe}
\end{equation}
The equation above is again an approximation valid if $\beta_2 << 1$.
We cannot obtain $P_e$ in a CW measurement.  However, 
using \cref{eq:qp} and
\cref{eq:qi}, we can express $P_e$ in terms of $P_t$:
\begin{equation}
P_e = \frac{\Qp}{\Qi} P_t\, .\label{eq:pe:pt}
\end{equation}
We can now substitute \cref{eq:pe:pt} into
\cref{eq:bpe} to obtain
\begin{equation}
\beta_1 \approx \frac{1}{2 \sqrt{\frac{\Qi}{\Qp} \frac{P_f}{P_t}} - 1}\, .\label{eq:bpt}
\end{equation}
The above equation can be easily seen to be equivalent to \cref{eq:qzeropt}.
Thus, consideration of the emitted power allows us to derive the same
results as obtained in the previous section, but using fewer steps.

The present approach provides additional insight if we return to the
definition of $\rto$ from \cref{eq:define} and use 
\cref{eq:s21} and \cref{eq:pe:pt}
to express $\rto$ in terms of $P_f$ and $P_e$:
\begin{equation}
\rto = \sqrt{\frac{P_e}{P_f}}\, .\label{eq:altrto}
\end{equation}
Thus the normalized scattering parameter relates the emitted wave
to the forward wave in the same way as $\sto$ relates the transmitted wave
to the forward wave.

The emitted power can be measured directly by modulation of the drive
amplitude.  However, mismatches in the input line may produce
systematic errors in the $P_e$ measurement, and the measurement
requires more specialized equipment (such as a spectrum analyzer or
oscilloscope and RF diode) and hence may require additional
calibration steps (as well as attention to the time resolution of the
measurement).  In effect, this analysis of the CW measurements makes
$P_t$ a proxy for $P_e$.

\subsection{Indirect Method: Additional Formulae,
Comments}

Using \cref{eq:qo} and \cref{eq:qp}, we can eliminate
$Q_0$ from \cref{eq:qzeropt} and express $P_d$ explicitly in terms of
$P_f$, $P_t$, and the coupling strengths:
\begin{equation}
P_d \approx P_t \frac{\Qp}{\Qi} \left(2 \sqrt{\frac{\Qi}{\Qp}\cdot\frac{P_f}{P_t}} - 1\right)\, .
\end{equation} 
As expected, we no longer need $P_r$ to calculate $Q_0$ and $P_d$, but we
now need $\Qi$.
For an alternative expression, using \cref{eq:define} and \cref{eq:s21}, we can express $P_d$ in terms of $\rto$ and $P_f$:
\begin{equation}
P_d \approx \rto \left(2 - \rto\right) P_f\, .
\end{equation}

Advantages of the indirect method include
(i) it does not use $P_r$, so systematic errors in the $P_r$
measurement are avoided;
(ii) it does not rely on the calculation of $P_f - P_r$, the
calculation of which may be problematic when not close to
unity coupling.

Disadvantages of the indirect method include
(i) the derivation of the formula is more complicated and its
application may be less intuitive;
(ii) it requires both $\Qi$ and $\Qp$;
(iii) the formula may not work well when the cavity is strongly
  overcoupled.  We will return to the latter point in \cref{S:lims}.

An additional disadvantage of the indirect method as derived in this
section is that we assume a weak pickup coupler.  The case of
arbitrary pickup coupler strength is discussed in \ref{S:CWarbInd}.
  
\section{Graphical Assessment Method: Derivation\label{S:DualTri}}

The graphical method to assess measurement consistency described in this report is used both for ``real
time'' assessments during SRF cavity cold tests at MSU
and for consistency evaluation in the final analysis of cold tests.

\cref{eq:betaoS} gives $\beta_1$ in terms of $\soo$;
\cref{eq:betaoR} gives  $\beta_1$ in terms of $\rto$.
Eliminating $\beta_1$ gives us
\begin{equation}
\frac{1\pm\soo}{1\mp\soo} = \frac{\rto}{2 - \rto}\, .\label{eq:DualRat}
\end{equation}
Solving for $\rto$ gives
\begin{equation}
\rto = 1 \pm \soo\, .\label{eq:duality}
\end{equation}

\begin{figure}[b]
\includegraphics[width=\columnwidth]{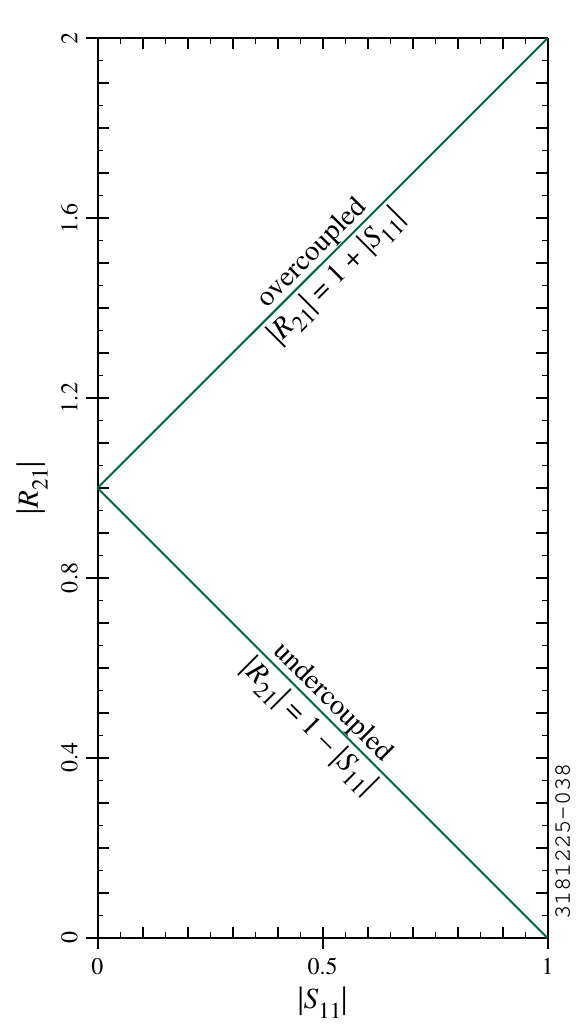}

\caption{The ``Duality Triangle.''  In CW measurements, the calculated
values of $\soo$ and $\rto$ should fall on the lower or upper green line
segments.\label{F:dual}}
\end{figure}

Since we must always have $0\leq\soo\leq 1$, 
\cref{eq:duality} tells us that the values of $\soo$ and $\rto$
should always be on one of the two equal sides of an isosceles
triangle, as shown in \cref{F:dual}.  We refer to it as the
``Duality Triangle.''  When doing CW measurements, $\soo$ and $\rto$
can be calculated from $\Qi$, $\Qp$, $P_f$, $P_r$, and $P_t$.
The proximity of these values to the Duality Triangle then provides an
indication of the self-consistency of the measured values.

As discussed, in \cref{S:emit}, the indirect analysis
effectively uses the transmitted wave of the pickup coupler as a proxy
for the emitted wave of the input coupler, since the latter cannot be
measured in CW\@; the vertical axis of the Duality Triangle can hence
be thought of as representing the scattering parameter for the emitted
wave, as inferred from the transmitted
wave.  If we use \cref{eq:altrto} and 
\cref{eq:s11} to rewrite \cref{eq:duality} in terms of
$P_f$, $P_r$, and $P_e$, we obtain
\begin{equation}
\sqrt{\frac{P_e}{P_f}} = 1 \pm \sqrt{\frac{P_r}{P_f}}\, .\label{eq:dualt}
\end{equation}
As a consistency check, we note that the above equation is equivalent
to Equation~(8.46) in Ref.~\cite{PKH1998}.

As discussed previously, the lower sign corresponds to the
undercoupled case, so that, from \cref{eq:duality}, $\rto$
is less than 1; for the upper sign, the overcoupled case, $\rto$ is
greater than 1.  Hence the lower side of the Duality Triangle
corresponds to the undercoupled case and the upper side corresponds to
the overcoupled case, as indicated in \cref{F:dual}.

Although we have assumed a weak pickup coupler for this derivation,
\ref{S:CWarbDual} shows that the results presented in this section are
valid for arbitrary pickup coupling strength, provided $\Qi$ and $\Qp$
are calculated properly (per \ref{S:ModArb}).  Thus, applications of
the Duality Triangle are not limited to the case of $\beta_2 \ll 1$,
and we replaced the ``approximately equal to'' signs with ``equal to''
signs in the equations above.

Though the Duality Triangle provides a way to check the
self-consistency of the measured values of $P_f$, $P_r$, and $P_t$
(used to obtain $\soo$ and $\sto$) with the calculated values of $\Qi$
and $\Qp$ (used to calculate $\rto$ from $\sto$), consistency with the
Duality Triangle does not guarantee that the results are valid.  One
counterexample is a scenario in which the calculated values of both
$\Qi$ and $\Qp$ are too large by a factor of 2.  In this case, $\rto$
is unaffected, and the Duality Triangle does not show a discrepancy.
Hence, as always, the cavity test analysis team should proceed with
care.

\section{Graphical Assessment Method: Vertices and Limiting Cases\label{S:lims}}

\subsection{Weak Input Coupling}
From \cref{eq:RtoB}, if $\beta_1 \ll 1$,
$\rto \approx 2\beta_1$; hence $\rto$ approaches zero for
$\beta_1 \rightarrow 0$.  Thus the lower vertex of the Duality
Triangle ($\soo = 1, \rto = 0$) corresponds to the weak input coupling
limit, $\beta_1 \ll 1$, or, equivalently, $\Qi \gg Q_0$.  In the
vicinity of the weak coupling limit, the $Q_0$ value from the direct
method may have excessive error due to $P_f - P_r$ being a small
difference between 2 large numbers; the $Q_0$ value from the indirect
method may be less prone to systematic error if the coupling strengths are
known with reasonable accuracy.  In this case, $P_e$ is small compared
to $P_f$ and $P_d$ and $P_r$ is approximately equal to $P_f$.

Explicit expressions from the
indirect method in the weak input coupling limit are
\begin{align}
Q_0 &\approx \frac{\rto}{2} = \frac{1}{2} \sqrt{\frac{P_t}{P_f}} \sqrt{\Qi \Qp}\, ,\label{eq:Qweak}\\
P_d &\approx 2\rto P_f = 2 \sqrt{P_t P_f} \sqrt{\frac{\Qp}{\Qi}}\, .
\end{align}
A variant of \cref{eq:Qweak} may be familiar to some
readers in the context of room temperature measurements on SRF
cavities.

\subsection{Unity Input Coupling}

Setting $\beta_1 = 1$ in \cref{eq:RtoB} gives $\rto = 1$.
Thus the middle vertex of the Duality Triangle ($\soo = 0, \rto = 1$)
corresponds to the unity input coupling case, $\beta_1 = 1$ and
$\Qi = Q_0$.  In the vicinity of unity coupling, one can hope that
both the direct method and the indirect method will produce reasonable
results.  Near unity coupling, $P_e$ is approximately equal to $P_f$ and
$P_r$ is approximately zero.

\subsection{Strong Input Coupling\label{S:strong}}

For large values of $\beta_1$, \cref{eq:RtoB} shows that $\rto$
approaches 2.  Thus, the upper vertex of the Duality Triangle
($\soo = 1, \rto = 2$) corresponds to the strong input coupling limit,
$\beta_1 \gg 1$ and $\Qi \ll Q_0$.  In the vicinity of the strong
coupling limit, the $Q_0$ value from the direct method may have
excessive error due to $P_f - P_r$ being a small difference between 2
large numbers; the $Q_0$ value from the indirect method may be
problematic as well, because the denominator in \cref{eq:qor}
approaches zero in the limit in which $\rto$ approaches 2.  The
measured power ratios are mainly determined by the couplers and are
not sensitive to the cavity $Q_0$.  In this case, $P_r$ is again
approximately equal to $P_f$ and $P_e$ is approximately equal to
$4 P_f$.  The latter result may be familiar to readers who use pulsed
systems.

\section{Detuning}

So far, we have assumed that the cavity is driven on resonance.  In this
section, we will consider the more general case in which the drive
frequency is not necessarily equal to the resonant frequency.

\subsection{Graphical Assessment Method: Detuning Predictions}

\begin{figure}[b]
\includegraphics[width=\columnwidth]{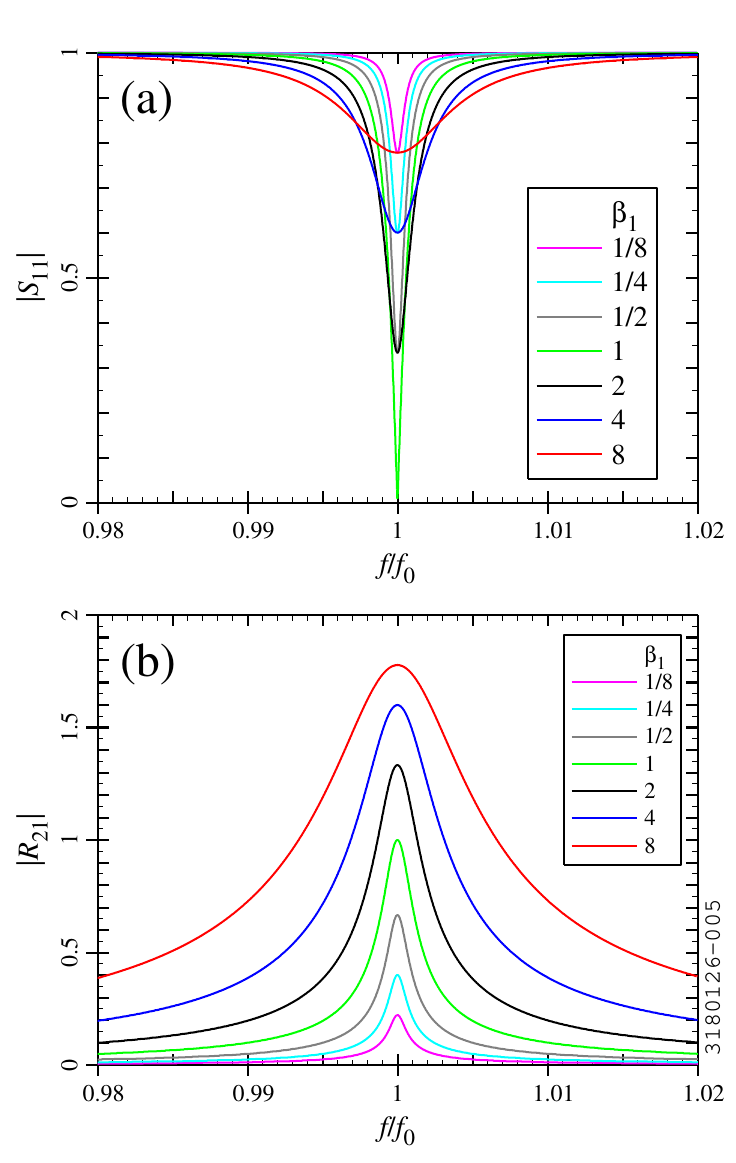}

\caption{Calculated scattering parameters as a function of normalized drive
  frequency for a cavity with $Q_0 = 10^3$: (a) $\soo$ and (b) $\rto$.\label{F:SnRvf}}
\end{figure}

Analyses of cavity behavior as a function of drive frequency can be
found in the literature, for example in Chapter 8 of
Ref.~~\cite{PKH1998}.  Useful formulae for the S-parameters as a
function of the drive angular frequency $\omega = 2 \pi f$ are
\begin{equation}
  S_{11} (\omega) = %
  \frac{\beta_1 - 1 - i Q_0 \left(\frac{\omega}{\omega_0} - \frac{\omega_0}{\omega}\right)}%
  {\beta_1 + 1 + i Q_0 \left(\frac{\omega}{\omega_0} - \frac{\omega_0}{\omega}\right)}\label{eq:soof}
\end{equation}
and
\begin{equation}
  |S_{21} (\omega)| = \sqrt{\frac{\Qi}{\Qp}} %
  \frac{2 \beta_1}{\sqrt{(\beta_1 + 1)^2 +
      \left[Q_0\left(\frac{\omega}{\omega_0} - \frac{\omega_0}{\omega}\right)\right]^2}} \, ,\label{eq:stof}
\end{equation}
where, again, $\omega_0$ is the resonant angular frequency.
The above are obtained from Eq.~(8.42) and Eq.~(8.30) in
Ref.~\cite{PKH1998}.  Using the definition of $\rto$, we can write

\begin{equation}
  |R_{21} (\omega)| = 
  \frac{2 \beta_1}{\sqrt{(\beta_1 + 1)^2 +
      \left[Q_0\left(\frac{\omega}{\omega_0} - \frac{\omega_0}{\omega}\right)\right]^2}}\, .
\end{equation}
The basic response of the cavity according to the above equations is
illustrated in \cref{F:SnRvf}, which shows $\soo$ and $\rto$ as a
function of the relative frequency ($f$ = drive frequency, $f_0$ =
resonant frequency) for $Q_0 = 10^3$ and $\beta_1$ values ranging from
0.125 to 8.
The patterns seen in \cref{F:SnRvf} are likely familiar to
most readers: there is a minimum in $\soo$ and a maximum in $\rto$ on
resonance; $\soo$ approaches 1 and $\rto$ approaches 0 for large
detuning; $\soo$ reaches 0 on resonance when $\beta_1 = 1$; the
bandwidth is smaller for low $\beta_1$ values and larger for high
$\beta_1$ values.  The maximum value of $\rto$ approaches 2 for large
$\beta_1$, consistent with \cref{S:strong}.

\begin{figure*}
\centering
\includegraphics[width=0.9\columnwidth]{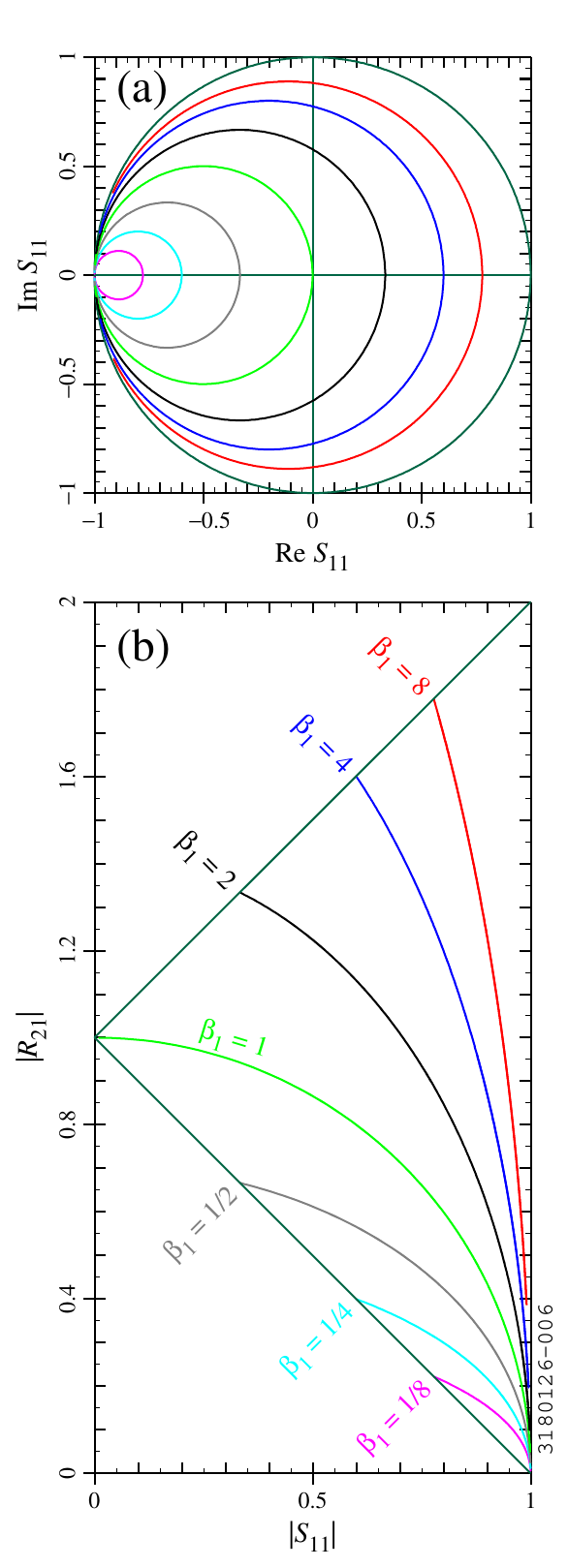}
\includegraphics[width=0.9\columnwidth]{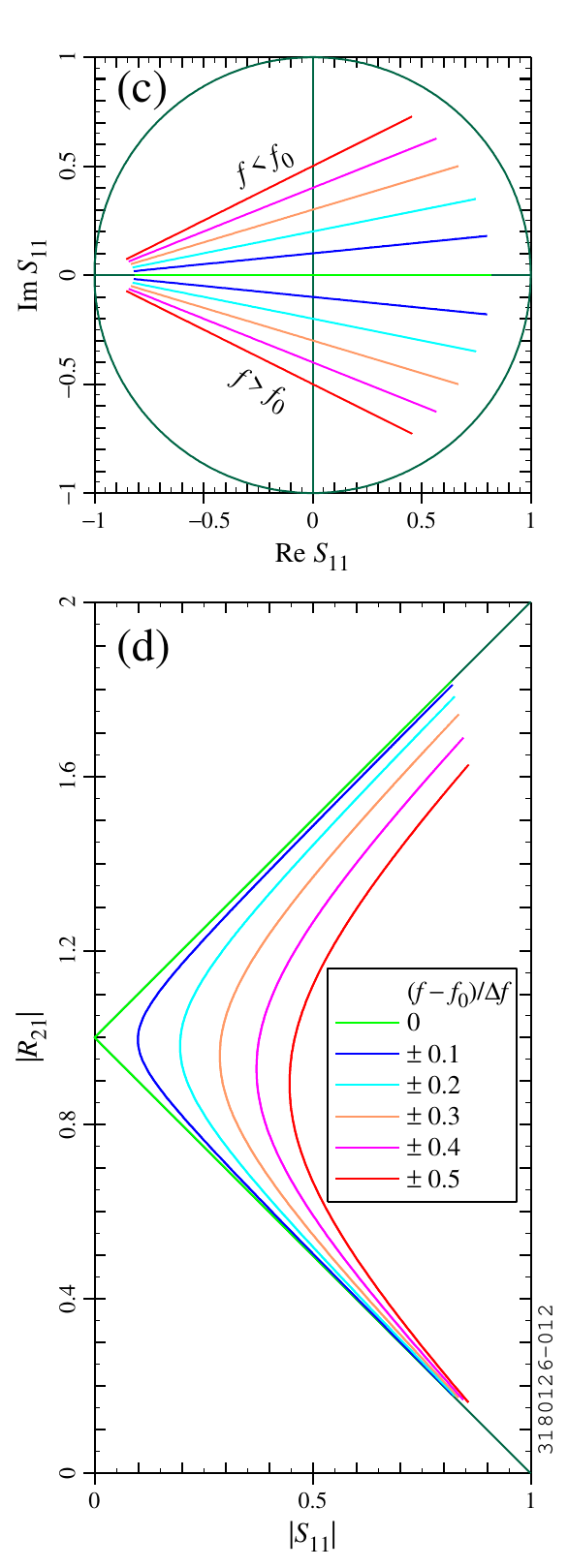}

\caption{(a) Polar plot of $S_{11}$ and (b) curves traced out in
  the $\soo$-$\rto$ plane as a function of drive frequency. (c) Polar
  plot of $S_{11}$ and (d) curves traces out in the $\soo$-$\rto$
  plane as a function of $\beta_1$.\label{F:PolTrivf}}
\end{figure*}

\cref{F:PolTrivf}a shows a polar plot of $S_{11}$ as a function of
drive frequency for different values of $\beta_1$, which will
likely be familiar to most reader as well.  As the drive frequency
varies, $S_{11}$ traces out a circle in the complex plane.  The path
traced out in the complex plane is dependent on $\beta_1$, but not on
$Q_0$.  The circle encloses the origin if $\beta_1 > 1$ but not if
$\beta_1 < 1$.

\cref{F:PolTrivf}b shows the corresponding curves traced out in the
$\soo$-$\rto$ plane as the drive frequency varies, with the Duality
Triangle shown in dark green.  Again, the path is dependent on
$\beta_1$, but not on $Q_0$.  On resonance, the values lie on the
Duality Triangle.  Off resonance points lie inside the Duality
Triangle.  This is consistent with \cref{F:SnRvf}a, in which the
minimum in $\soo$ occurs on resonance, and $\soo$ increases when we go
off resonance.  As we go further and further off resonance, we
approach a ``fully detuned'' limit, $\soo =1$, $\rto = 0$, which is
the same as the weak coupling limit.

\cref{F:PolTrivf}c shows a polar plot of $S_{11}$ as a function of
$\beta_1$ for for different values of relative drive frequency offset.
The on-resonance case is a horizontal line through the origin (green).
The off-resonance cases produce straight lines that do not intersect
the origin.  Each curve had a different value of $(f - f_0)/\Delta f$;
normalizing to the loaded bandwidth $\Delta f$ should produces curves
similar to the those obtained by driving the cavity in phase-lock mode
with a constant offset in the loop phase.  (The results which be
somewhat different for a an open-loop scenario with the cavity drive
with a constant offset in the drive frequency.)

\cref{F:PolTrivf}d shows the corresponding curves traced out in the
$\soo$-$\rto$ plane as $\beta_1$ varies, with the Duality Triangle
shown in dark green.  On resonance, the values lie on the Duality
Triangle (light green).  Off-resonance curves lie inside the Duality
Triangle, as expected.  Positive and negative relative frequency
offsets lie on the same curve in the $\soo$-$\rto$ plane.  Once again,
the paths traced out in the complex plane and in the $\soo$-$\rto$
plane are independent of $Q_0$.

\subsection{Graphical Assessment Method: Detuning Measurements\label{S:TriDetM}}

\begin{figure*}
\centering
\includegraphics[width=0.9\columnwidth]{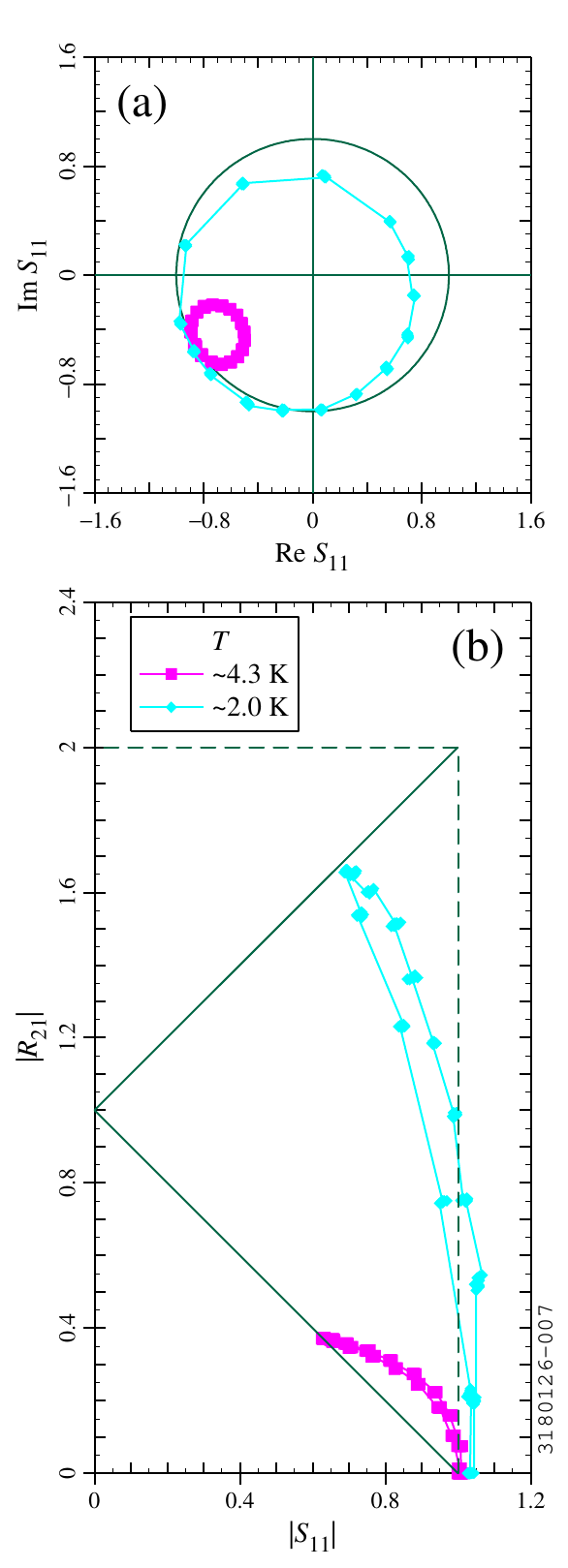} %
\includegraphics[width=0.9\columnwidth]{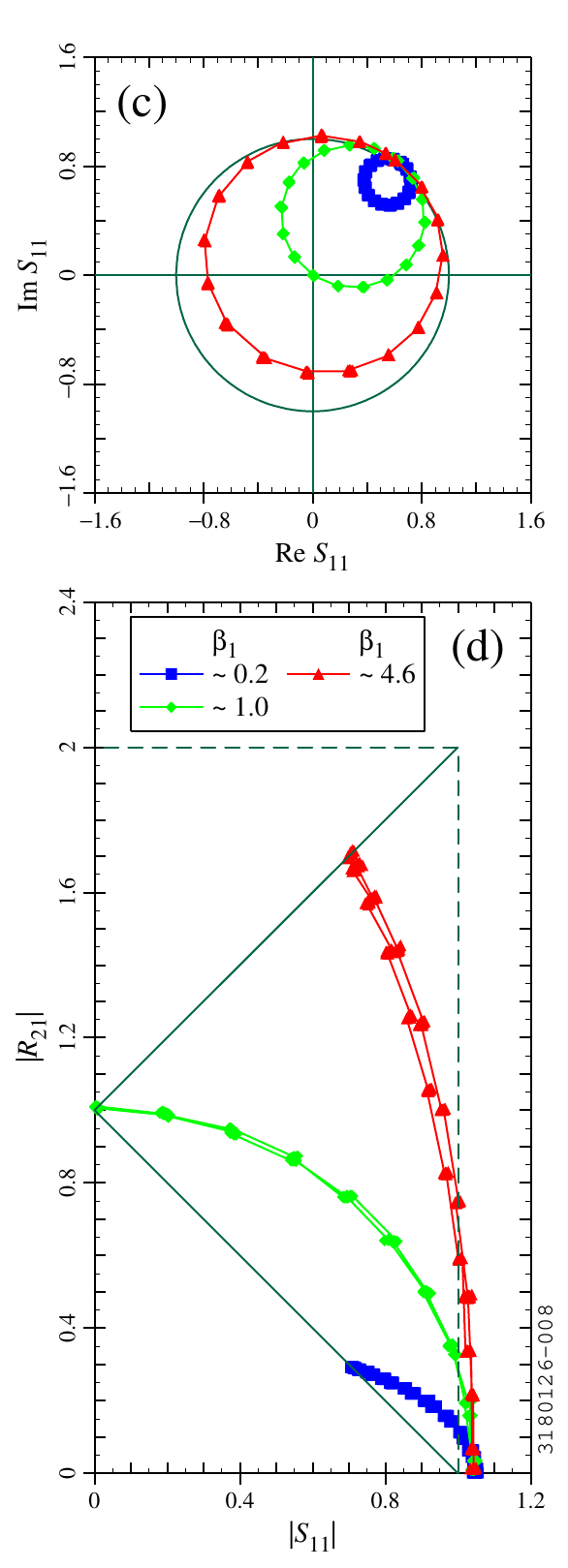} %

\caption{CW measurements on FRIB HWRs as a function of loop phase: (a) polar plot of $S_{11}$ 
  and (b) $\soo$ vs $\rto$
  at 2 different bath temperatures (S53-028, December 2017); (c) polar
  plot of $S_{11}$ and (d)  $\soo$ vs $\rto$ for 3 different coupler
  positions (S53-121, October 2018).\label{F:PolTriHWRs}}
\end{figure*}

In the case of CW measurements on SRF cavities, it is impractical to
sweep the drive frequency, as the bandwidth is typically comparable to
the resonant frequency fluctuations from microphonics.  However, in
phase lock-loop mode, a similar effect can be achieved by varying the
loop phase.  \cref{F:PolTriHWRs} shows some examples of such
measurements on FRIB HWRs.  \cref{F:PolTriHWRs}a and
\cref{F:PolTriHWRs}c show polar plots for different $\beta_1$ values.
In these measurements, we set the forward power to establish a low
field in the cavity ($E_a$ between 1 and 2 MV/m typically) and kept
the forward power constant which varying the loop phase, so that the
cavity field decreased as the loop phase offset increased.  The
results are consistent with expectations, though the curves are
rotated by an arbitrary phase offset relative to the predictions of
\cref{F:PolTrivf}a (this is as expected, as the measured phases are
not corrected for the phase advance along the transmission lines
between the couplers and the phase measurement planes); In addition,
due to systematic errors, $\soo$ is slightly larger than 1 for some
cases.

\cref{F:PolTriHWRs}b and \cref{F:PolTriHWRs}d show the corresponding
results in the $\soo$-$\rto$ plane.  Again, the results are consistent
with expectations, although $\soo$ is again slightly larger than 1 for
some cases.  As predicted, the values lie on the Duality Triangle when
on resonance; the values lie inside the Triangle when off resonance;
and the values approach the lower right vertex for large detuning.

In \cref{F:PolTriHWRs}a and \cref{F:PolTriHWRs}b, the coupler position
was fixed and bath temperature was adjusted to change from undercoupled
to overcoupled.  The same values of $\Qi$ and $\Qp$ are used in the
data analysis (obtained via modulated measurements).

In \cref{F:PolTriHWRs}c and \cref{F:PolTriHWRs}d, the bath temperature
was fixed (4.3~K), but a variable coupler was used, allowing us to
achieve different $\beta_1$ values via coupler position adjustment.
The values of $\Qi$ were not measured for every coupler position, so
estimated $\Qi$ values were used, based on the measured $S_{11}$
values, the measured $Q_0$ values, and consistency of the Duality
Triangle.

Per \cref{F:PolTrivf}b, we expect negative and positive detuning to
take us along the same path in the plane, but the measurements of
\cref{F:PolTriHWRs}b and \cref{F:PolTriHWRs}d show some discrepancies
between positive and negative detuning, as seen by the double-valued
curves.  This tends to be more pronounced in the overcoupled cases.
The cause may again be systematic errors in the measured values.

\subsection{Quality Factor Calculation with Detuning: Predictions}

Ideally, CW measurements are done with the drive frequency ($f$) equal
to the resonant frequency ($f_0$).  In practice, there may be some
offset between $f$ and $f_0$ due to systematic errors, imperfect drive
frequency tuning in the open-loop case, or imperfect optimization of
the loop phase in the phase-lock case.  Hence we cannot always be sure
that $Q_0$ is calculated with $f$ exactly equal to $f_0$.

When using the direct method, the calculated $Q_0$ depends on $\soo^2$
and $\sto^2$.  Using \cref{eq:QdirS}, \cref{eq:soof}, and
\cref{eq:stof}, we can see that, at least in the case of a weak pickup
coupler, the calculated $Q_0$ is in principle the same for $f \neq f_0$
and $f = f_0$.  In this case, both $\sto^2$ and $1 - \soo^2$ depend on
the frequency, but their ratio does not.

\begin{figure}[tb]
\includegraphics[width=\columnwidth]{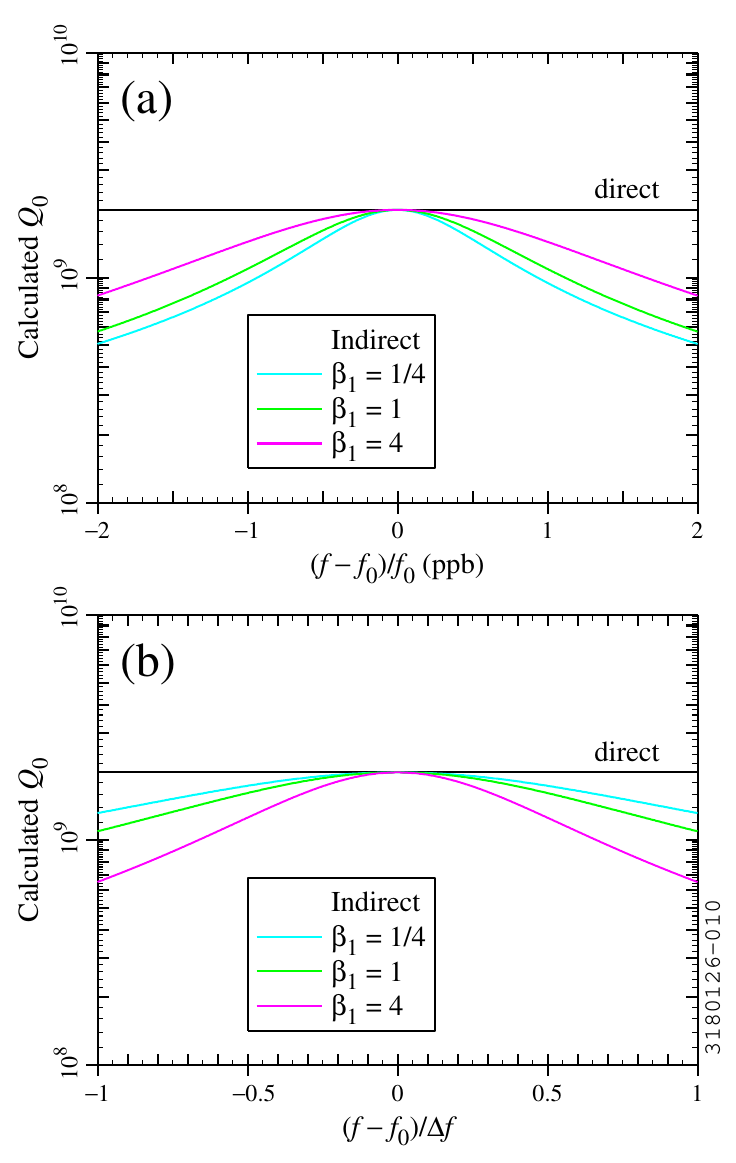} %

\caption{Calculated $Q_0$ values via the
  indirect method as a function of (a) relative detuning in
  parts per billion and (b) detuning normalized to the bandwidth
  $\Delta f$, with an actual $Q_0$ of $2 \tten{9}$; black line: $Q_0$
  calculated via the direct method.\label{F:PredQ0vf}}
\end{figure}

This is in contrast to the indirect method, in which $Q_0$ depends on
$\rto$, per \cref{eq:qor}.  As $\rto$ changes with frequency, the
calculated $Q_0$ follows suit.  These findings are illustrated in
\cref{F:PredQ0vf}a: the calculated $Q_0$ is independent of detuning
when we use the direct method (black), but decreases with detuning
when we use the indirect method (light colors).  The calculated
indirect $Q_0$ decreases more rapidly with detuning when undercoupled
(cyan) and decreases more slowly with detuning when overcoupled
(magenta).  These trends are qualitatively consistent with the loaded
bandwidth being smaller when undercoupled and larger when overcoupled.

\begin{figure}
\includegraphics[width=\columnwidth]{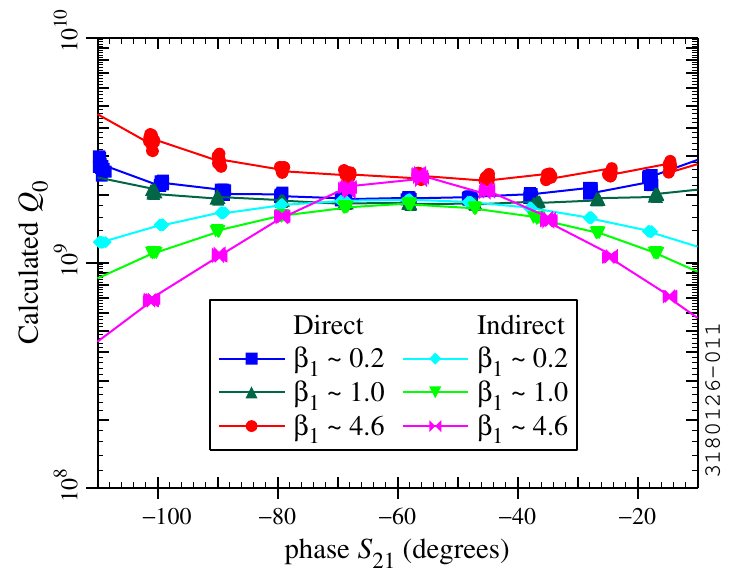} %

\caption{Calculated $Q_0$ values via the direct method (dark colors)
  and indirect method (light colors) as a function of loop phase based
  on CW measurements on a FRIB HWR (S53-121, October 2018) at 4.3 K.\label{F:MeasQ0vf}}
\end{figure}

\cref{F:PredQ0vf}b shows the same results, but with the frequency
normalized to the loaded bandwidth $\Delta f$ defined via
\cref{eq:BW}.  This normalization should be on par with the case of a
loop phase offset in phase-lock mode.  With this normalization, the
overcoupled case shows greater sensitivity to detuning than the
undercoupled case.  In other words, with the indirect method, detuning
by the same fraction of the bandwidth or by the same loop phase offset
should produce a bigger $Q_0$ error with stronger coupling.

\subsection{Quality Factor Calculation with Detuning: Measurements}

We can calculate $Q_0$ as a function of detuning for the measurements
as a function of loop phase of \cref{S:TriDetM}.  \cref{F:MeasQ0vf}
shows the corresponding $Q_0$ calculations for the results shown in
\cref{F:PolTriHWRs}c and \cref{F:PolTriHWRs}d (measurements at 4.3~K,
3 different coupler positions).  As described above, the measurements
were done at a relatively low accelerating gradient
($E_a \sim 1.5$~MV/m when on resonance).  The direct method (dark
colors) shows a small increase in calculated $Q_0$ for large detuning.
This may be explained in part by the decrease in the cavity field with
detuning (with the forward power fixed while varying the loop phase),
which should produce a slight increase in the measured $Q_0$ due to
``$Q$-slope'' (an example of the $Q$-slope at 4.3 K can be seen in
\cref{F:Qdir:alt3} below).  We estimate that the $Q$-slope should
produce an increase in $Q_0$ of about 10\%, which is less than the
worst-case differences seen in the direct-calculation $Q_0$ values.
Hence there may be other effects contributing to the apparent $Q_0$
increase.

The indirect method (light colors) shows more pronounced decreases in
the calculated $Q_0$ values with detuning.  As discussed above,
adjustment of the loop phase should be analogous to shifting the drive
frequency relative to the loaded bandwidth, so we would expect the
behavior to be similar to that seen in \cref{F:PredQ0vf}b.  The
predictions of \cref{F:PredQ0vf}b and the measurements of
\cref{F:MeasQ0vf} are indeed consistent, at least at qualitatively.

Ideally, the calculated $Q_0$ values on resonance should be the same
for the undercoupled, matched, and overcoupled cases.  However,
\cref{F:MeasQ0vf} shows that the calculated $Q_0$ is a bit higher
for the overcoupled case.  This is true for both the direct method
(red circles) and the indirect method (magenta bowties).  This may be
due to systematic errors in the measurements.  Though we did not
measure $\Qi$ for each coupler position, the consistency between the
indirect and direct calculation suggests that the difference in
calculated $Q_0$ is not due to an error in $\Qi$.

Predictions and measurements show that the indirect method is at a
disadvantage relative to the direct method, as the calculated $Q_0$ is
more sensitive to detuning.  However, this should not be a problem if
the cavity can be driven reasonably close to the resonant frequency.
The frequency sensitivity of the indirect method is nevertheless
something to keep in mind for scenarios in which the frequency offset
is inadvertently high.

\section{Applications and Examples}

All FRIB production cavities were Dewar tested after jacketing
\cite{NIMA1014:165675}.  In a typical test, CW and modulated
measurements are first done at 4.3~K, along with conditioning of
multipacting barriers if needed.  The cavity is then cooled to 2~K by
pumping on the helium bath; low-field CW measurements are usually done
during the pump-down to infer the low-field $Q_0$ as a function of
temperature.  CW and modulated measurements are repeated at 2~K\@.  The
cool-down to 2~K is often paused at 3~K for additional modulated
and/or CW measurements.  Most of the Dewar tests are done with FRIB
low-level RF controllers \cite{NAPAC2019:WEPLM03}, which allow for
both amplitude and phase measurements of the forward, reverse, and
transmitted signals.  The majority of the tests have been done with
fixed RF couplers, though some tests were done with variable input
couplers (allowing us to adjust the coupler position during the cold
test).  Cold tests for the FRIB energy upgrade are done with an analog
RF system without phase measurements.

The Duality Triangle has been found to be useful for real-time
feedback during Dewar certification tests.  We have used it to
trouble-shoot issues with the RF system and check the consistency of
field level calibrations.

All FRIB production cryomodules were cold tested prior to installation
into the linac tunnel \cite{NAPAC2019:WEPLM73}.  The Duality Triangle
allows for real-time feedback during cryomodule tests.  Some examples
of Duality Triangle applications for Dewar tests and cryomodule tests
will be presented in this section.

\subsection{Dewar Tests: Real-Time}

\begin{figure}
\centering
\includegraphics[width=0.9\columnwidth]{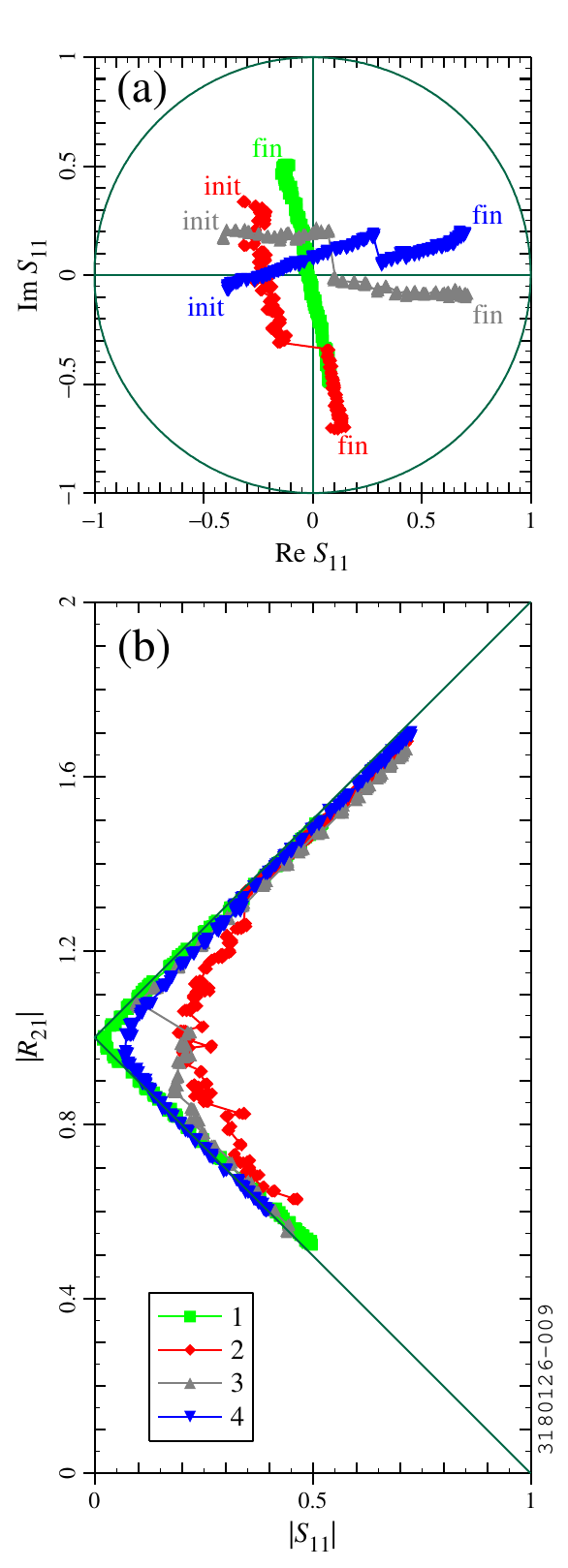}

\caption{CW measurements on FRIB HWRs while pumping from 4.3~K to 2~K: (a) polar plot of $S_{11}$ 
  and (b) $\soo$ vs $\rto$ (S53-159 and S53-161, total of 4 tests, May
  2023 to Jul 2024).\label{F:PolTriHWRrt}}
\end{figure}

One application of the Duality Triangle is to detect loop phase errors
in phase-lock loop mode.  Some examples are included in
\cref{F:PolTriHWRrt}, which shows CW measurements taken on FRIB
$\beta_m = 0.54$ HWRs while pumping on the liquid helium bath to
reduce the temperature from 4.3~K to 2~K ($\beta_m$ = optimum
normalized beam speed $v/c$).  We expect $Q_0$ to increase as the
temperature increases, such that we should go from undercoupled at
4.3~K to overcoupled at 2~K\@.  In the first data set (green squares),
we observe that the measurements of $S_{11}$ fall along a straight
line that intersects the origin (\cref{F:PolTriHWRrt}a) and the data
points fall on the sides of the Duality Triangle
(\cref{F:PolTriHWRrt}b).  This is the expected behavior and is
consistent with the green curve of \cref{F:PolTrivf}c and
\cref{F:PolTrivf}d (taking into consideration that $S_{11}$ has a
phase offset, so the on-resonance line is not necessarily horizontal).

For the other data sets (red diamonds, gray and blue triangles),
initially the $S_{11}$ curve does not intersect the origin in
\cref{F:PolTriHWRrt}a (``init'' denotes the initial undercoupled
measurements) ; and the data do not intersect the left vertex of the
Duality Triangle in \cref{F:PolTriHWRrt}b.  Per \cref{F:PolTrivf}c and
\cref{F:PolTrivf}d, both of these observations point to the
possibility that the loop phase is such that the cavity is driven off
resonance (with a lesser offset in the case of the blue triangles and
with a greater offset in the case of the red diamonds and gray
triangles).  Partway through the pump-down, we adjusted the loop
phase.  As a result, for the upper portion of the Duality Triangle,
the data lie closer to the side of the Duality Triangle
(\cref{F:PolTriHWRrt}b), and a portion of the $S_{11}$ curve can be
extrapolated back to intersect the origin (\cref{F:PolTriHWRrt}a;
``fin'' denotes the final overcoupled measurements, after phase
adjustments).

The need to adjust the loop phase during the pump-down is not
unexpected.  As seen in \cref{{F:SnRvf}}a, the minimum in $S_{11}$ is
sharper when $\beta_1$ is close to 1, so the sensitivity to phase
errors is more acute when we are near the left vertex of the Duality
Triangle.  Likewise, per \cref{F:PolTrivf}d, a given loop phase error
produces a bigger discrepancy relative to the Duality Triangle near
the left vertex.

We note that, in the case of FRIB cavity tests, loop phase errors can
be seen via both the polar $S_{11}$ plot and the Duality Triangle.  On
the other hand, for more typical cases in which the RF phases are not
measured, a polar $S_{11}$ plot cannot be used to detect loop phase
errors, but the Duality Triangle can be.

\subsection{Dewar Tests: Final Analysis}

\begin{figure*}[tb]
\includegraphics[width=\textwidth]{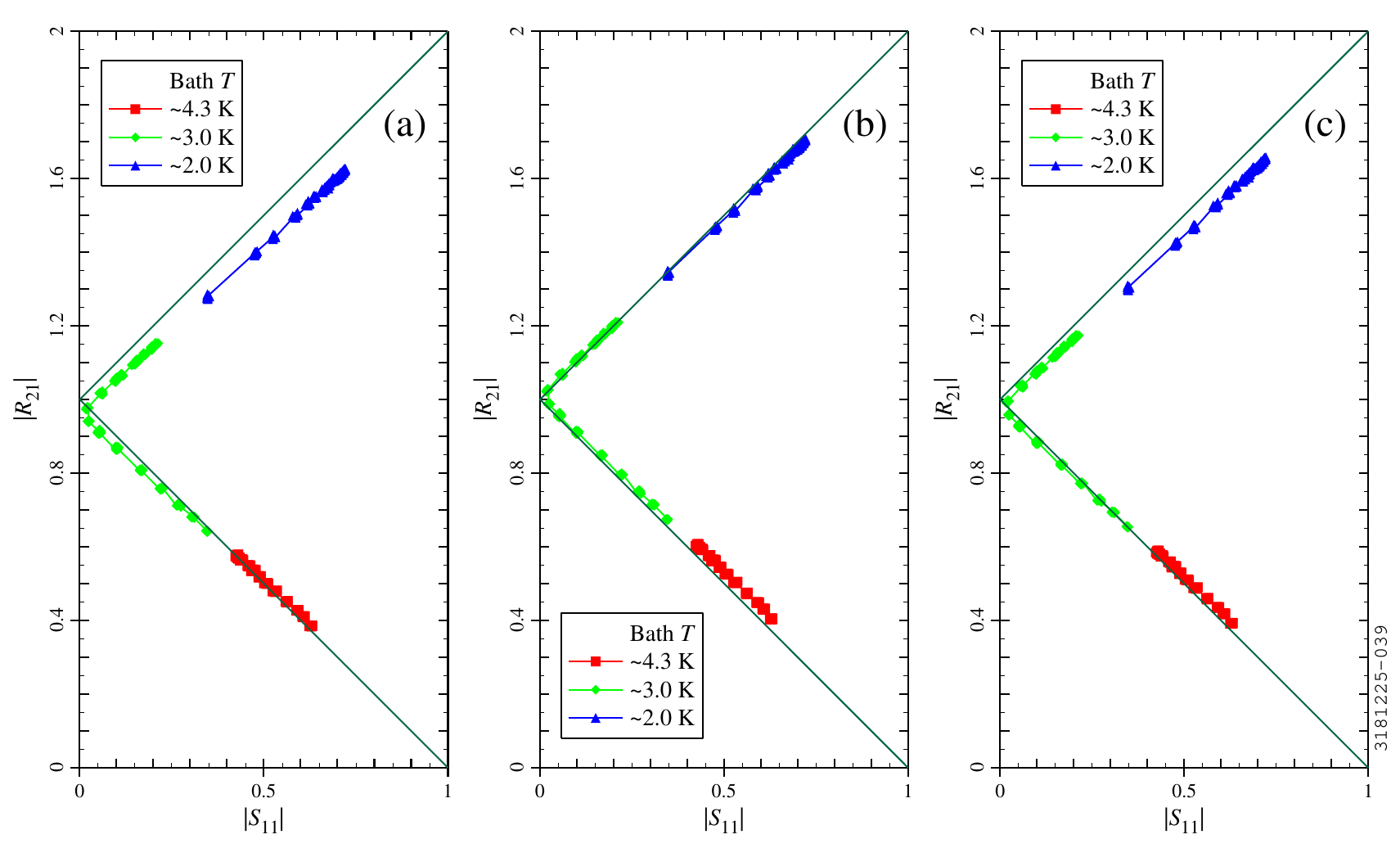}

\caption{Scattering parameters for CW measurements on a $\beta_m = 0.54$ HWR at three different
  bath temperatures.  The coupling strengths are obtained from
  modulated measurements at (a) $\sim 4.3$~K, (b) $\sim 3.0$~K, and (c) $\sim 2.0$~K.\label{F:dual53}}

\end{figure*}

In the final analysis stage, we have used the Duality Triangle to
assess the self-consistency of field level calibrations based on
modulated measurements at different temperatures.  Ideally, all of the
modulated measurements should give the same results for coupling
strengths, but we typically observe differences from one measurement
to another.  As an example, the coupling strengths calculated for a
FRIB $\beta_m = 0.54$ HWR (S53-155, a spare HWR tested in October
2022)
from modulated measurements at different temperatures ($T$ = bath
temperature) are shown in \cref{T:cpl}.  The spread in values is
about 10\% for both $\Qi$ and $\Qp$.

\begin{table}[b]
\caption{Calculated coupling strength values from modulated
  measurements on a $\beta_m = 0.54$ HWR at different
  temperatures.\label{T:cpl}}

\begin{center}
\begin{tabular}{ccc}\toprule
$T$ &       & \\
(K) & $\Qi$ & $\Qp$ \\ \midrule
  $\sim 4.3$ & $6.69\tten{9}$ & $5.60\tten{11}$ \\
  $\sim 3.0$ & $6.27\tten{9}$ & $5.79\tten{11}$ \\
  $\sim 2.0$ & $6.05\tten{9}$ & $5.26\tten{11}$ \\ \bottomrule
\end{tabular}
\end{center}

\end{table}

CW measurements were done at the same three bath temperatures.  The
corresponding values of $\soo$ and $\rto$ from the CW measurements are
shown in \cref{F:dual53}.  The measured values (markers) do not fall
exactly on the sides of the Duality Triangle (dark green lines).
Overall, the coupling strengths obtained at $\sim 3$~K provide the
best consistency (\cref{F:dual53}b), although there are discrepancies
for the undercoupled measurements (the red squares, particularly).
The coupling strengths obtained at $\sim 4.3$~K (\cref{F:dual53}a) and
$\sim 2.0$~K (\cref{F:dual53}c) clearly show more overall discrepancy,
disagreeing especially in the overcoupled cases (blue
triangles, particularly).  Note that the modulated measurements at
$\sim 3$~K had the least input coupler mismatch (closest to unity coupling).

\begin{figure}
\includegraphics[width=\columnwidth]{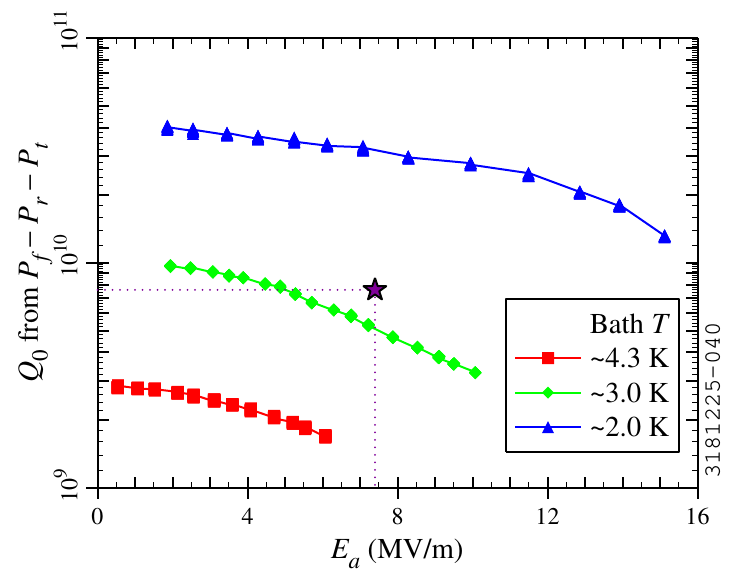}

\caption{Quality factor calculated via the direct method as a function of accelerating gradient for CW measurements on a $\beta_m = 0.54$ HWR at three different
  bath temperatures; the coupling strengths are obtained from
  modulated measurements $\sim 3.0$~K for all cases. Purple star:
  design goal for operation at 2~K.\label{F:Qdir:alt3}}

\end{figure}

During FRIB cavity production, we used the modulated measurement with
the best consistency for the final analysis of the CW measurements.
In this example, we used the modulated measurements at $\sim 3$~K\@.
Because the input coupler was approximately matched for typical FRIB
cavity Dewar tests, we used the direct method to calculate $Q_0$ in
the final analysis for all FRIB production cavities.  The
final-analysis values of $Q_0$ as a function of field for the present
example are shown in \cref{F:Qdir:alt3}.

\begin{figure}[tb]
\includegraphics[width=\columnwidth]{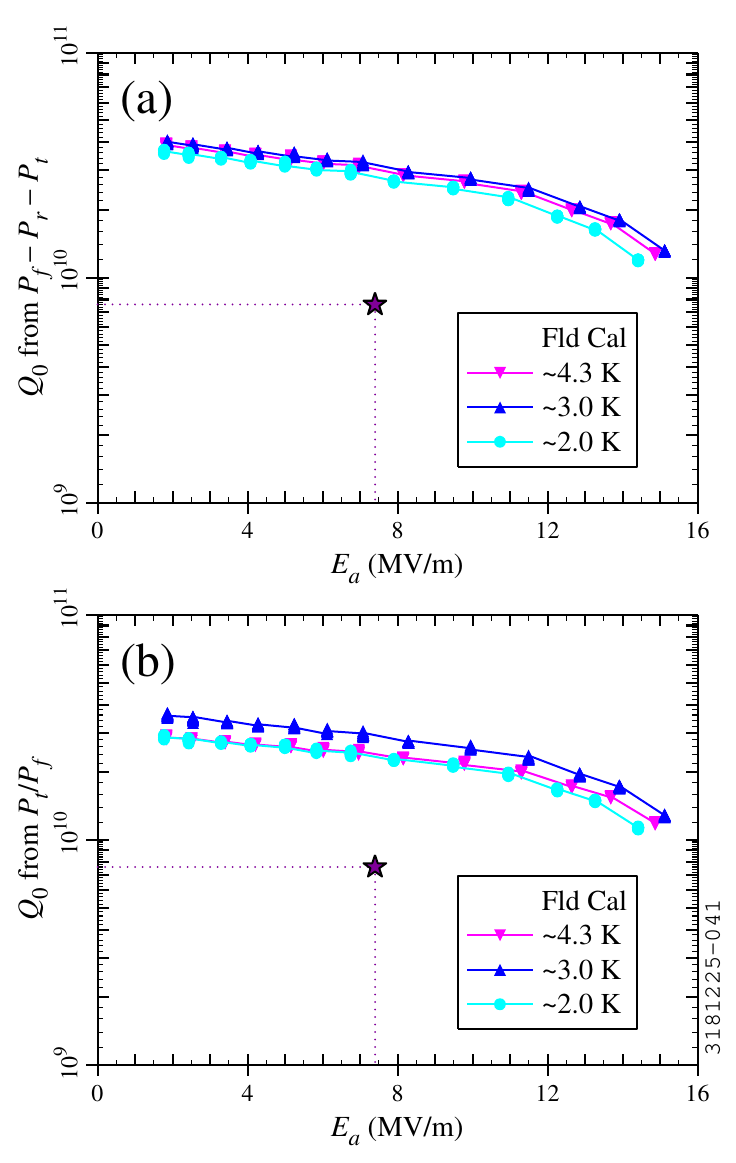}

\caption{Quality factor calculated via the (a) direct method and (b)
  the indirect method as a function of $E_a$ for CW measurements on a
  $\beta_m = 0.54$ HWR at $\sim 2$~K; the different curves correspond to
  the coupling strengths from
  modulated measurements at different temperatures.\label{F:Qdir:ind}}

\end{figure}

For reference, \cref{F:Qdir:ind} compares the direct and indirect methods to obtain
$Q_0$ from CW measurements at $\sim 2$~K and compares the results from the modulated
measurements at different temperatures.  In this example, which is
typical of FRIB cavity certification tests, the spread in $Q_0$ values
from one case to another is approximately 25\% or less;
the differences in $E_a$ from one case to another are approximately 8\%
or less.

The present example shows that systematic errors are not negligible
for FRIB resonator tests.  The Duality Triangle provides a method to
assess consistency and try to minimize discrepancies.  The
discrepancies suggest that a more advanced vector correction method
may be beneficial for reduction of systematic errors, as discussed in
\cref{S:intro}.  Although the Duality Triangle helps to recognize
discrepancies and minimize their adverse impact, a vector correction
method could in principle remove the systematic errors in a more
rigorous way and significantly improve the accuracy of the CW
measurement results.

\subsection{Cryomodule Tests}

\begin{figure*}[tb]
\includegraphics[width=\textwidth]{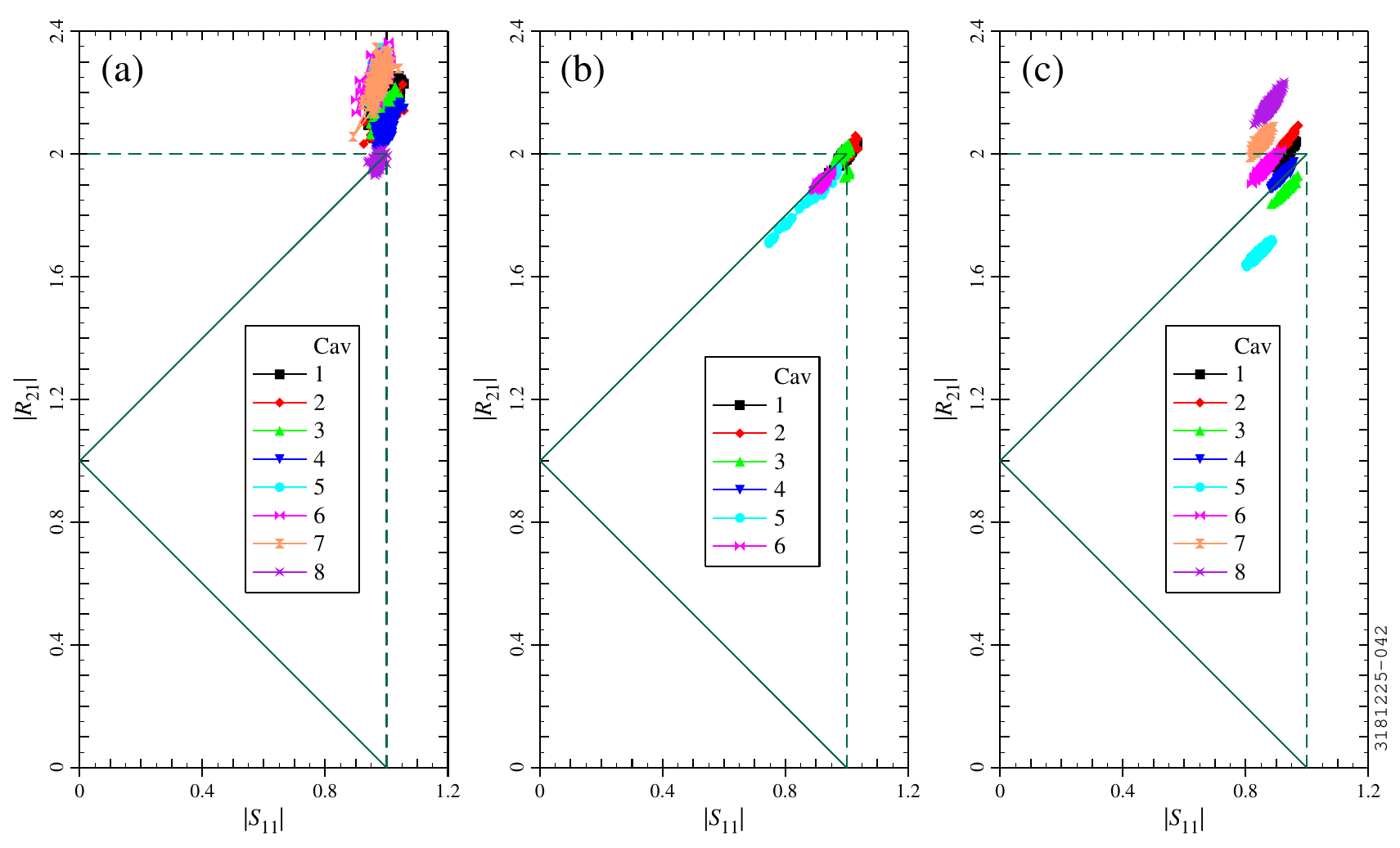}

\caption{Scattering parameters for CW measurements on FRIB
  cryomodules: (a) a $\beta_m = 0.086$ QWR cryomodule; (b) a
  $\beta_m = 0.29$ HWR cryomodule; (c) a $\beta_m = 0.54$ HWR
  cryomodule.\label{F:cryomod}}

\end{figure*}

Some examples of ``near-real-time'' cryomodule test results are shown
in \cref{F:cryomod}.  As expected, all of the measured values are near
the strong coupling limit ($\soo = 1, \rto = 2$), as the FPCs are set
to be overcoupled for control of amplitude and phase in the presence
of microphonics and beam loading.

\cref{F:cryomod}a shows results for a $\beta_m = 0.086$ cryomodule
containing 8 QWRs (SCM813, tested in July 2023).  Field emission
X-rays were observed for some of the the cavities at high field.
There is significant scatter between cavities, such that sometimes
$\soo > 1$ and $\rto > 2$; this confirms that the systematic errors in
the measured values of $P_f$, $P_r$, and $P_t$ are significant enough
to make it difficult to infer $P_d$ and $Q_0$.

\cref{F:cryomod}b shows results for a $\beta_m = 0.29$ HWR cryomodule
containing 6 HWRs (SCM207, tested in early February 2019).  No field emission was
observed, but some of the CW measurements were taken while
conditioning the high multipacting barrier.  The results are more
consistent with the Duality Triangle, but some of the values still
have $\soo > 1$ and $\rto > 2$, again indicating that the systematic
errors are significant.

\cref{F:cryomod}c shows results for a $\beta_m = 0.54$ HWR cryomodule
containing 8 HWRs (SCM508, tested in late February 2019).  In this
case, some field emission X-rays were observed for 2 out of 8
cavities.  Again, some of the CW measurements were taken while
conditioning the high multipacting barrier.  These results show more
scatter than the $\beta_m = 0.29$ example, with less overlap between
cavities than for the $\beta_m = 0.086$ example.  Some of the cavities
have $\rto > 2$, but, fortuitously, none have $\soo > 1$.

It is clear from \cref{F:cryomod} that, even in the most consistent
cases, the systematic errors are large enough to make it impractical
to obtain $P_d$ and $Q_0$ via the direct method.  Moreover, as
discussed in \cref{S:strong}, the indirect method is impractical in
the strong input coupling case.  As a result, $P_d$ and $Q_0$ must be
inferred from calorimetric measurements rather than RF power
measurements for FRIB cryomodules, a situation which is not atypical
for SRF cryomodules.

\section{Conclusion}

The case of continuous-wave measurements on a two-port radio-frequency
cavity is considered.  The assumption of a weak pickup coupler allows
for simplifications, but the derivations can be generalized to the
case of an arbitrary pickup coupling strength.  The quality factor of
the cavity can be calculated via both direct and indirect methods, the
latter being possible if both coupling strengths are known.  The
indirect method is useful in the weak input coupling case when
systematic errors are present.  Graphical methods allow for rapid
evaluation of the consistency of the measurements and coupling
strengths, which has been found useful for both real-time feedback
during cavity tests and subsequent analysis of cavity test results.

\section*{Acknowledgments}
\addcontentsline{toc}{section}{Acknowledgments}

We thank colleagues at NSCL and FRIB who have been part of the MSU SRF
team over the past 25 years.  FRIB SRF cavity testing is a
collaborative effort between the cavity fabrication team, the cavity
surface preparation team, and the FRIB cryogenics team, with support
from the FRIB radio-frequency team.  Thank you to David Meidlinger for
typesetting the original hand-written derivations included in
\cref{S:back}, \cref{S:dir}, and \cref{S:ind} and sharing the source
file.

This material is based upon work supported by the U.S. Department of
Energy, Office of Science, Office of Nuclear Physics, and used
resources of the Facility for Rare Isotope Beams (FRIB) Operations,
which is a DOE Office of Science User Facility under Cooperative
Agreement DE-SC0023633.  Additional support was provided by the State
of Michigan and Michigan State University.

\appendix

\section{Analysis of Modulated Measurements\label{S:Mod}}

Descriptions of modulated measurements can be found in a number of
textbooks, for example, Sections 8.3-8.4 of Ref.~\cite{PKH1998}.  In
this appendix, we first provide an overview of modulated measurements
and then proceed with the assumption of a weakly-coupled pickup, which
is usually a safe assumption, particularly for SRF cavities; this is
an assumption made in Ref.~\cite{PKH1998}.  We consider a more general
analysis valid for arbitrary pickup coupling strength in
\ref{S:ModArb}.

In the case of NRF cavities, frequency domain measurements are
typically done in lieu of the modulated measurements (time domain)
described in this section.  However, the time domain discussion can be
applied to the frequency domain by replacing the
time-domain decay time measurement with a frequency-domain loaded
bandwidth measurement; determination of the coupling in the frequency
domain can be done using an $S_{11}$ polar plot in lieu of the
transient behavior of $P_r$.

\subsection{Modulated Measurement Basics\label{S:ModBase}}

In contrast to the case of CW measurements with known input and pickup
coupling strengths, the case of modulated measurements with unknown
input and pickup coupling strengths is not overdetermined.  Hence
modulated measurements do not allow for multiple methods of
calculating $Q_0$; likewise, values of $\soo$ and $\rto$ obtained from
a modulated measurement will always lie on the Duality Triangle.

The goal of a modulated measurement is to obtain $Q_0$, $\Qi$, $\Qp$,
and $U$ from measured quantities.  In a typical modulated measurement,
we first measure the usual CW quantities, typically at low field:
$P_f$, $P_r$, $P_t$, and $f_0$.  Then we turn off or modulate the
drive power to obtain the decay time $\tau_L$ from the rate of
decrease in $P_t$ and calculate the loaded quality factor $Q_L$ from
$\tau_L$.  We determine whether the cavity-input-coupler pair is
undercoupled or overcoupled by comparing the reverse power upon
turning on the drive power ($P_r = P_f$ at turn-on) and turning off
the drive power ($P_r = P_e$ at turn-off): undercoupled if
$P_e < P_f$, overcoupled if $P_e > P_f$.

We can calculate the CW value of $P_d$ via \cref{eq:pdiff}; we can
then obtain $\beta_2$ via \cref{eq:b2p}.  The steps so far do not
require any assumptions about weak pickup coupling, but the
calculation of $\beta_1$ must be done with additional care in the case
of arbitrary pickup strength; we discuss the calculation of $\beta_1$
in \ref{S:ModWeak} for the weak pickup case and \ref{S:ModArb} for the
general case.
Once we have calculated $\beta_1$, we can obtain $Q_0$ from $Q_L$ by
recasting \cref{eq:QLinv} as
\begin{equation}
Q_0 = Q_L \left(1 + \beta_1 + \beta_2\right)\, .
\end{equation}
With $Q_0$, $\beta_1$, and $\beta_2$ being known, we can obtain $\Qi$
and $\Qp$ from \cref{eq:b1} and \cref{eq:b2}.  Finally, we
can obtain $U$ via \cref{eq:qo}.

\subsection{Weak Pickup Coupling\label{S:ModWeak}}

With the assumption that $\beta_2 \ll 1$, we can obtain $\beta_1$ from
$P_f$ and $P_r$ using \cref{eq:betaRat}.  This allows us to calculate
$Q_0$, $\Qi$, $\Qp$, and $U$ in a straightforward way via the steps
outlined in the previous section.

\subsection{Arbitrary Pickup Coupling Strength\label{S:ModArb}}

In the case in which we cannot assume a weakly-coupled pickup, the
cavity-input-coupler match may be affected by the pickup coupler.  Let
us consider the cavity and pickup coupler as a ``subsystem'' with
quality factor $\Qos$.  We write
\begin{equation}
\frac{1}{\Qos} = \frac{1}{Q_0} + \frac{1}{\Qp}\, .\label{eq:Qos}
\end{equation}
Furthermore, let us define an ``input-coupler-to-subsystem'' coupling
factor $\beta_{1s}$ via
\begin{equation}
\beta_{1s} \equiv \frac{\Qos}{\Qi}\, .
\end{equation}
Making use of \cref{eq:b2}, we can rewrite \cref{eq:Qos} as
\begin{equation}
\frac{Q_0}{\Qos} = 1 + \beta_2\, .
\end{equation}
Using the definitions of $\beta_1$ and $\beta_{1s}$ and the equation above, we can write
\begin{equation}
\beta_1 = \frac{Q_0}{\Qi} = \frac{\Qos}{\Qi} \frac{Q_0}{\Qos} = \beta_{1s} \left(1 + \beta_2\right)\, .\label{eq:b1b1s}
\end{equation}
The case of a 2-port cavity with arbitrary coupling is considered
briefly in Section 5.5 of Wangler \cite{WANGLER2008}; one finding is that
the condition for a matched input coupler is $\beta_1 = 1 + \beta_2$.
Our result of $\beta_1 = \beta_{1s} (1 + \beta_2)$ from the above
equation is consistent with Wangler's result, as it implies that the
matched input condition is $\beta_{1s} = 1$.

Using \cref{eq:b2p}, we can replace $\beta_2$ with $P_t$
(measured) and
$P_d$ (calculated from $P_f$, $P_r$, and $P_t$):
\begin{equation}
\beta_1 = \beta_{1s} \left(1 + \frac{P_t}{P_d}\right)\, .
\end{equation}
Considering the input coupler as being coupled to the cavity-pickup subsystem
rather than just to the cavity, we rewrite \cref{eq:betaRat} as follows:
\begin{equation}
\beta_{1s} = \frac{1 \pm \sqrt{P_r/P_f}}{1 \mp \sqrt{P_r/P_f}}\, .\label{eq:betaStg}
\end{equation}
Using the last 2 equations, we can eliminate $\beta_{1s}$ to obtain an
expression for $\beta_1$:
\begin{equation}
\beta_1 = \left(1 + \frac{P_t}{P_d}\right) \frac{1 \pm \sqrt{P_r/P_f}}{1 \mp \sqrt{P_r/P_f}}\, .
\end{equation}
In terms of measured quantities, we can write
\begin{equation}
\beta_1 = \left(\frac{P_f - P_r}{P_f - P_r - P_t}\right) \frac{1 \pm \sqrt{P_r/P_f}}{1 \mp \sqrt{P_r/P_f}}\, .
\end{equation}
The equation above can be seen to be similar to \cref{eq:betaRat},
with the distinction that we do not assume a weak pickup coupler.  It
is clear that it reduces to \cref{eq:betaRat} in the limit
$P_t \ll P_d$.  We can use this equation to calculate $Q_0$, $\Qi$,
$\Qp$, and $U$ for an arbitrary pickup coupler strength via the steps
outlined in \ref{S:ModBase}.

\section{Analysis of CW Measurements with Arbitrary Pickup Coupling Strength\label{S:CWarb}}

In this appendix, we consider the calculation of $Q_0$ and the
application of the Duality Triangle for cases which do not satisfy
$\beta_2 \ll 1$.  We conclude by applying these analysis techniques to
the case of a cavity test in which the input and pickup coupling
strengths are approximately equal.

\subsection{Direct Method}

The direct method (\cref{S:dir}) does not require any assumptions
about $\beta_2$.  Hence \cref{eq:Qdir} and \cref{eq:QdirS} are valid
for arbitrary pickup coupling strength.  We still need $\Qp$ to
calculate $U$, so \ref{S:ModArb} is applicable when using the direct
method with arbitrary $\beta_2$.

\subsection{Indirect Method\label{S:CWarbInd}}

We will use the ``emitted wave'' approach of \cref{S:emit} for this
derivation.  The same results can be obtained via the ``brute force''
approach of \cref{S:ind}.

Following the ``subsystem'' approach of \ref{S:ModArb}, we formulate a
generalized version of \cref{eq:bpe} as
\begin{equation}
\beta_{1s} = \frac{1}{2 \sqrt{\frac{P_f}{P_e}} - 1}\, .\label{eq:b1s:emit}
\end{equation}
We can eliminate $\beta_{1s}$ using \cref{eq:b1b1s} to obtain
\begin{equation}
\beta_1 = \frac{1 + \beta_2}{2 \sqrt{\frac{P_f}{P_e}} - 1}\, .
\end{equation}
Using \cref{eq:b1} and \cref{eq:b2} to eliminate $\beta_1$ and $\beta_2$, and using \cref{eq:altrto} to
eliminate $P_f/P_e$ we obtain
\begin{equation}
\frac{Q_0}{\Qi} = \frac{1 + \frac{Q_0}{\Qp}}{\frac{2}{\rto} - 1}\, .
\end{equation}
Solving for $Q_0$, we obtain
\begin{equation}
Q_0 = \frac{\Qi}{\frac{2}{\rto} - 1 - \frac{\Qi}{\Qp}}\, .
\end{equation}
The above equation is valid for arbitrary $\beta_2$ values.  It
is clear that it reduces to \cref{eq:qor} in the limit $\Qp \gg \Qi$.
An explicit expression for $Q_0$ in terms of measured powers and
coupling strengths is
\begin{equation}
Q_0 = \frac{\Qi}{2\sqrt{\frac{\Qi}{\Qp}\cdot\frac{P_f}{P_t}} - 1 - \frac{\Qi}{\Qp}}\, .
\end{equation} 
Likewise, an expression for $P_d$ in terms of measured powers and
coupling strengths is
\begin{equation}
P_d = P_t \frac{\Qp}{\Qi} \left(2 \sqrt{\frac{\Qi}{\Qp}\cdot\frac{P_f}{P_t}} - 1 - \frac{\Qi}{\Qp}\right)\, ,
\end{equation} 
and a corresponding expression in terms of $\rto$ and $P_f$ is
\begin{equation}
P_d = \rto \left[2 - \rto\left(1 + \frac{\Qi}{\Qp}\right)\right] P_f\, .
\end{equation}

\subsection{Graphical Assessment Method\label{S:CWarbDual}}

The Duality Triangle derivation of \cref{{S:DualTri}} made use of
\cref{eq:betaoS} and \cref{eq:betaoR}, both equations for $\beta_1$.
Analogous equations valid for arbitrary $\beta_2$ are \cref{eq:betaStg},
which can be rewritten as
\begin{equation}
\beta_{1s} = \frac{1\pm\soo}{1\mp\soo}
\end{equation}
and \cref{eq:b1s:emit}, which can be rewritten as
\begin{equation}
\beta_{1s} = \frac{1}{\frac{2}{\rto}-1} = \frac{\rto}{2 - \rto}\, .
\end{equation}
Elimination of $\beta_{1s}$ once again leads us to \cref{eq:DualRat}, which we
can see to be valid for arbitrary $\beta_2$.  Hence, provided $\Qi$
and $\Qp$ are calculated correctly (per \ref{S:Mod}), the Duality Triangle can be
applied for arbitrary pickup coupling strength.

\subsection{Example: Cold Test with Approximately Equal Input and
  Pickup Coupling Strengths}

A recent cold test with a pickup coupler of coupling strength similar
to that of the input coupler can serve to illustrate the analysis
methods described above.  The coupling strengths calculated from
modulated measurements at about 2~K are shown in \cref{T:QextWeak}.
Given that $Q_0$ is of order $1 \tten{10}$ near 2~K, it is clear that
the usual assumption of $\beta_2 \ll 1$ cannot be justified.  The
assumption of weak pickup coupling leads to significantly different
values for the coupling strengths relative to the values obtained with
the more general approach of \ref{S:ModArb}.

\begin{table}[htb]
\caption{Calculated coupling strength values from modulated
  measurements on a $\beta_m = 0.65$ cavity with different assumptions.\label{T:QextWeak}}

\begin{center}
\begin{tabular}{lcc}\toprule
Pickup coupling strength & $\Qi$ & $\Qp$ \\ \midrule
Assume weak & $4.10\tten{10}$ & $1.72\tten{10}$ \\
Arbitrary & $2.26\tten{10}$ & $2.12\tten{10}$ \\ \bottomrule
\end{tabular}
\end{center}

\end{table}

\begin{figure*}[tb]
\centering
\includegraphics[width=\columnwidth]{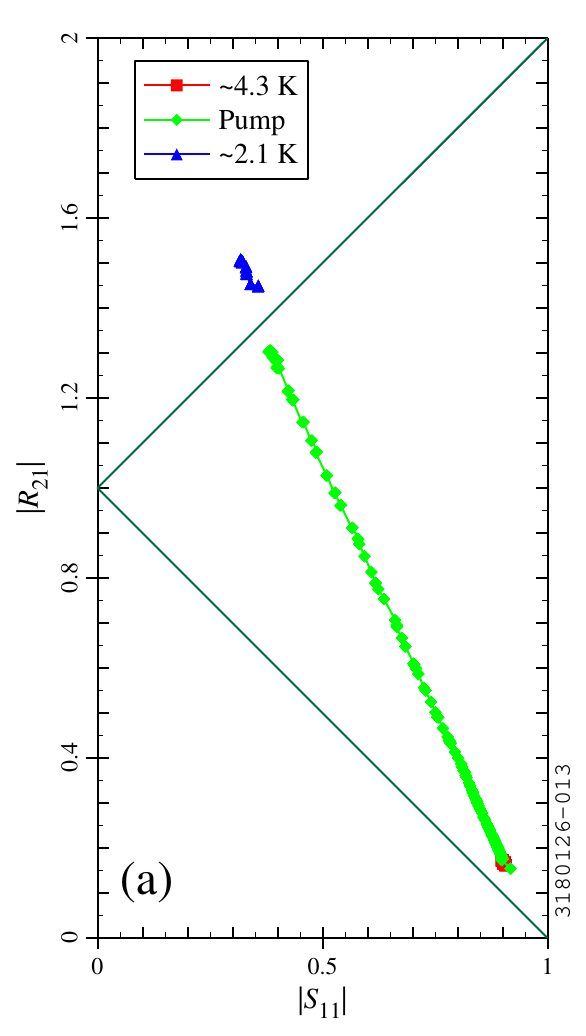}
\includegraphics[width=\columnwidth]{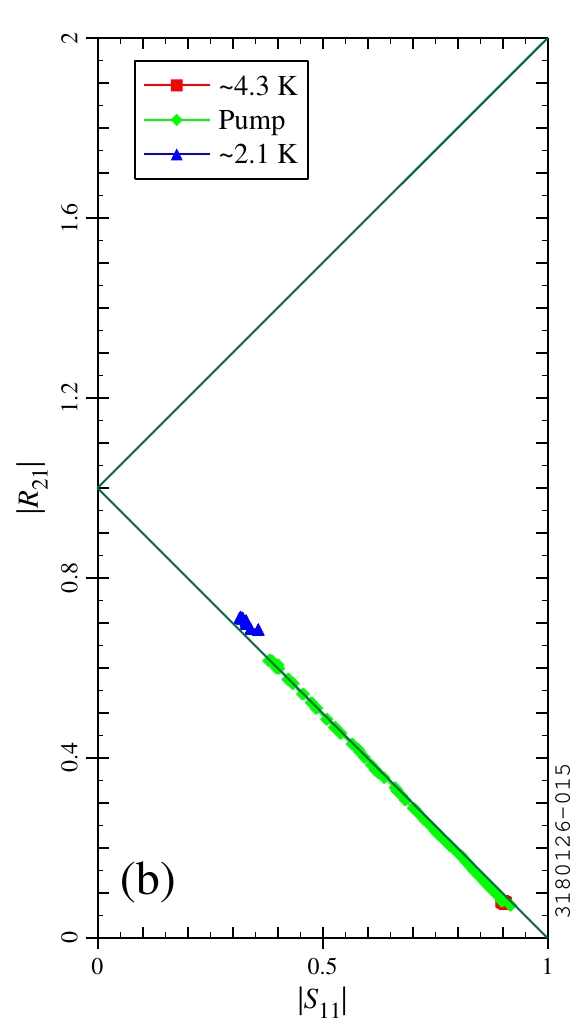}

\caption{Duality Triangle for a cold test on a cavity with
  approximately matched pickup coupling (S65-004, tested in October
  2024). Coupling strengths are obtained from modulated measurement
  (a) assuming or (b) not assuming a weak pickup.\label{F:WeakTri}}

\end{figure*}

The discrepancy can be seen clearly via the Duality Triangle, as
illustrated in \cref{F:WeakTri}, which shows CW measurements taken
while pumping on the bath to reduce the temperature from 4.3~K to 2~K
(green diamonds) as well as some measurements before and after the
pump-down (red squares, blue triangles).  When the coupling strengths
are calculated with the weak pickup assumption (per \ref{S:ModWeak}), the measured CW values
do not fall on the sides of the Duality Triangle (\cref{F:WeakTri}a),
indicating that the calculated value of $\Qp/\Qi$ is inconsistent with
the measured values of $P_f$, $P_r$, and $P_t$.  When the coupling
strengths are calculated without assuming a weak pickup (per \ref{S:ModArb}), the
results are approximately consistent with the Duality Triangle, with
the measured values all having $\beta_1 < 1$ (\cref{F:WeakTri}b).

\begin{figure*}[tb]
\centering
\includegraphics[width=\columnwidth]{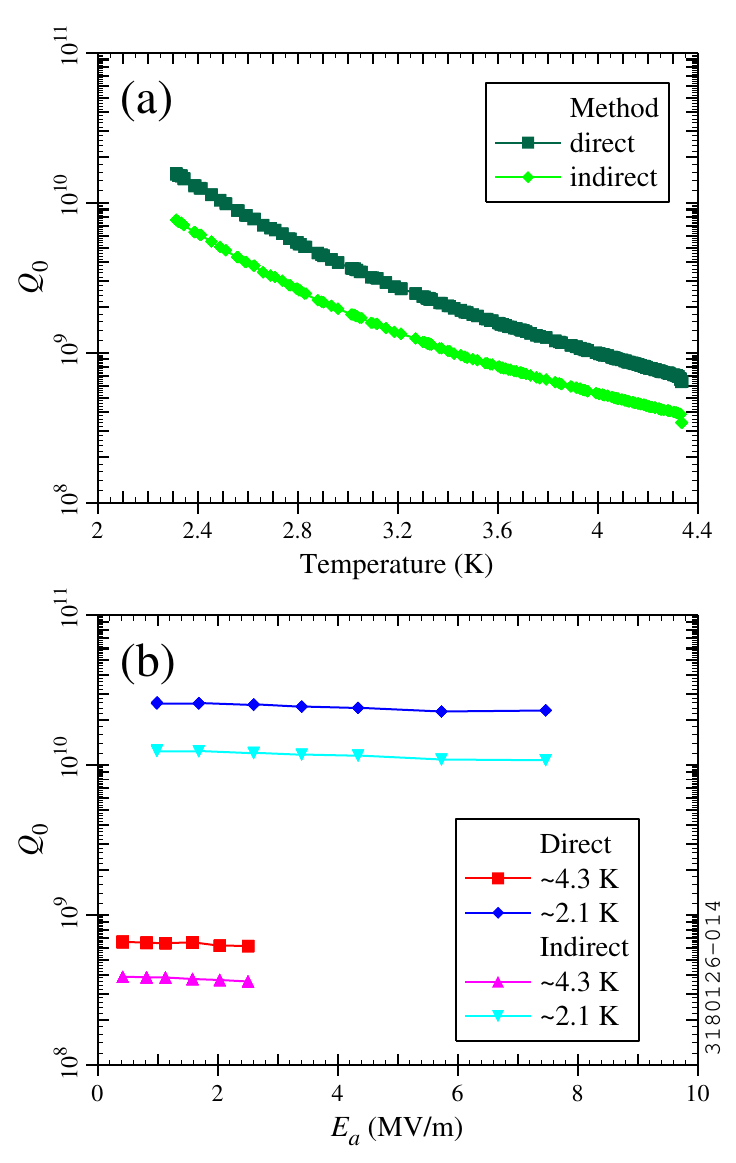}
\includegraphics[width=\columnwidth]{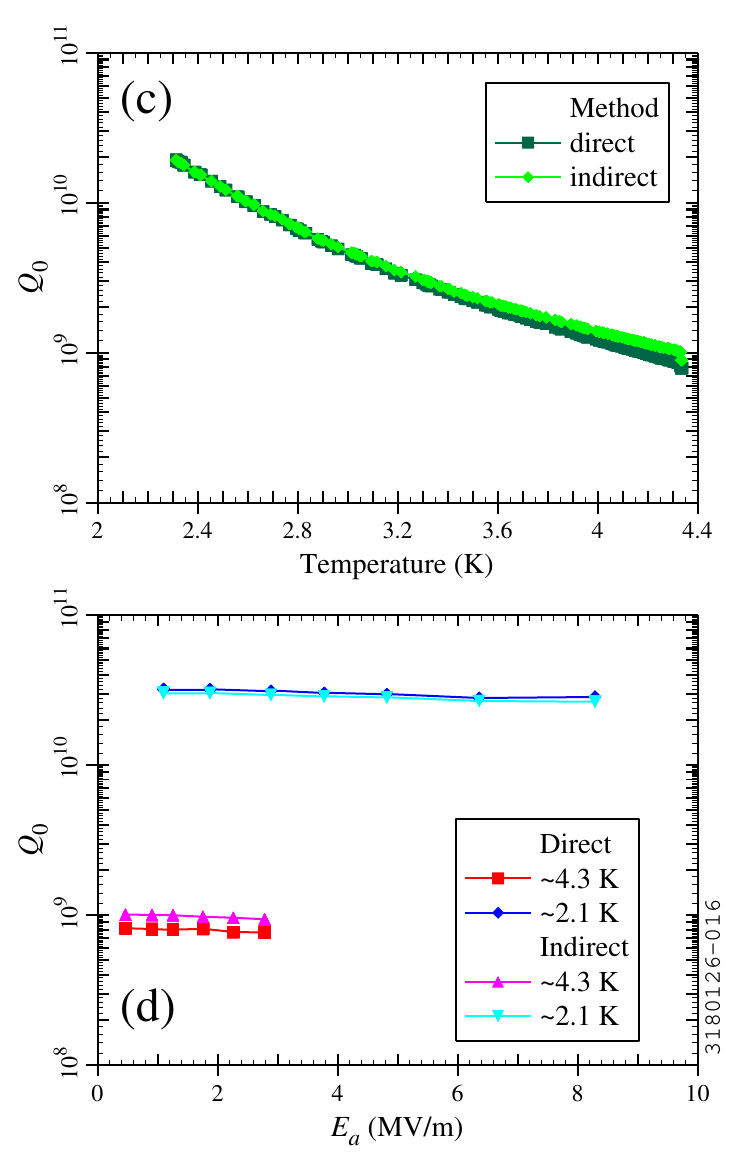}

\caption{Calculated quality factor as a function of (a, c) temperature
  and (b, d) accelerating gradient ($E_a$) for the cavity test with
  approximately matched pickup coupling.  Coupling strengths and
  indirect $Q_0$ values are calculated (a, b) assuming or (c, d) not
  assuming a weak pickup.\label{F:WeakQTE}}

\end{figure*}

Corresponding quality factors as a function of temperature and field
are shown in \cref{F:WeakQTE}.  Curves on the left-hand side assume
weak pickup coupling for both the calculation of the coupling
strengths and the indirect calculation of $Q_0$.  The discrepancy in
$Q_0$ values between the direct method (dark colors) and indirect
method (light colors) is significant.

Curves on the right-hand side of \cref{F:WeakQTE} allow for an
arbitrary pickup coupling strength for the analysis of the modulated
measurements and the indirect calculation of $Q_0$.  The agreement
between $Q_0$ values from the direct and indirect method is
significantly improved, though not perfect.  As seen in
\cref{F:WeakQTE}, wrongly assuming weak coupling causes us to infer a
significantly lower value for the indirect $Q_0$.  The weak pickup
coupling assumption mistakenly considers the power transferred to the
pickup coupler to be dissipated in the cavity, thereby mistakenly
finding a lower value of $Q_0$.

\bibliographystyle{elsarticle-num-names}
\bibliography{dual}

\end{document}